\documentclass[journal,comsoc]{IEEEtran}
\usepackage[sort,compress]{cite}
\usepackage[T1]{fontenc}
\usepackage{mathtools,dsfont,eucal}
\usepackage{graphicx,booktabs,wrapfig}
\makeatletter
\let\MYcaption\@makecaption
\makeatother
\usepackage[cmintegrals]{newtxmath}
\usepackage[font=footnotesize]{subcaption}
\usepackage{array,booktabs,ragged2e}

\makeatletter
\let\@makecaption\MYcaption
\makeatother

\usepackage[standard,amsmath,thmmarks]{ntheorem}
\usepackage[ruled,vlined]{algorithm2e}
\usepackage{hyperref}

\usepackage{microtype}

\newcommand{\IND}{\mathds{1}}
\newcommand{\PR}{\mathds{P}}
\newcommand{\EXP}{\mathds{E}}

\newcommand{\ACCEPT}{0}
\newcommand{\REJECT}{1}
\newcommand{\OFF}{\mathrm{off}}
\newcommand{\OV}{\mathrm{ov}}

\newcommand\GRAD{\nabla}
\DeclareMathOperator\PROJ{Proj}

\ifCLASSINFOpdf
\else
\fi
\usepackage[cmintegrals]{newtxmath}
\usepackage{array}
\usepackage{fixltx2e}

\usepackage{stfloats}
\usepackage{url}

\begin{document}
%
\title{Structure-aware reinforcement learning for node-overload protection in mobile edge computing}
%
%
%

\author{Anirudha~Jitani,~\IEEEmembership{Student Member,~IEEE,}
        Aditya~Mahajan,~\IEEEmembership{Senior Member,~IEEE,}
        Zhongwen~Zhu,~\IEEEmembership{Member,~IEEE,}
        Hatem~Abou-zeid,~\IEEEmembership{Member,~IEEE,}
        Emmanuel~Thepie~Fapi,~\IEEEmembership{}
        and~Hakimeh~Purmehdi,~\IEEEmembership{Member,~IEEE}
\IEEEcompsocitemizethanks{\IEEEcompsocthanksitem Anirudha Jitani is with the School of Computer Science, McGill University and Montreal Institute of Learning Algorithms. Aditya Mahajan is with the Department of Electrical and Computer Engineering, McGill University, Canada. \protect\\ Emails:
\href{mailto:anirudha.jitani@mail.mcgill.ca}{anirudha.jitani@mail.mcgill.ca}, \href{mailto:aditya.mahajan@mcgill.ca}{aditya.mahajan@mcgill.ca}
\IEEEcompsocthanksitem Zhongwen Zhu, Emmanuel Thepie Fapi, and Hakimeh Purmehdi are with Global AI Accelerator, Ericsson, Montreal, Canada. Hatem Abou-zeid is with Ericsson, Ottawa, Canada. \protect\\ Emails: 
\href{mailto:zhongwen.zhu@ericsson.com}{zhongwen.zhu@ericsson.com},
\href{mailto:emmanuel.thepie.fapi@ericsson.com}{emmanuel.thepie.fapi@ericsson.com},
\href{mailto:hakimeh.purmehdi@ericsson.com}{hakimeh.purmehdi@ericsson.com},
\href{mailto:hatem.abou-zeid@ericsson.com}{hatem.abou-zeid@ericsson.com}.
\IEEEcompsocthanksitem This work was supported in part by MITACS Accelerate, Grant IT16364.}}
\maketitle

\begin{abstract}

Mobile  Edge  Computing  (MEC)  refers  to  the  concept  of  placing  computational capability and applications at the edge of the network, providing benefits such as reduced latency in handling client requests, reduced network congestion, and improved performance of applications. The performance and reliability of MEC are degraded significantly when one or several edge servers in the cluster are overloaded. Especially when a server crashes due to the overload, it causes service failures in MEC. In this work, an adaptive admission control policy to prevent edge node from getting overloaded is presented. This approach is based on a recently-proposed low complexity RL (Reinforcement Learning) algorithm called SALMUT (Structure-Aware Learning for Multiple Thresholds), which exploits the structure of the optimal admission control policy in multi-class queues for an average-cost setting. We extend the framework to work for node overload-protection problem in a discounted-cost setting. The proposed solution is validated using several scenarios mimicking real-world deployments in two different settings --- computer simulations and a docker testbed. Our empirical evaluations show that the total discounted cost incurred by SALMUT is similar to state-of-the-art deep RL algorithms such as PPO (Proximal Policy Optimization) and A2C (Advantage Actor Critic) but requires an order of magnitude less time to train, outputs easily interpretable policy, and can be deployed in an online manner.

\end{abstract}


\section{Introduction}

\IEEEPARstart{I}{n} the last decade, we have seen a shift in the computing paradigm from co-located datacenters and compute servers to cloud computing. Due to the aggregation of resources, cloud computing can deliver elastic computing power and storage to customers without the overhead of setting up expensive datacenters and networking infrastructures. It has specially attracted small and medium-sized businesses who can leverage the cloud infrastructure with minimal setup costs. In recent years, the proliferation of Video-on-Demand (VoD) services, Internet-of-Things (IoT), real-time online gaming platforms, and Virtual Reality (VR) applications has lead to a strong focus on the quality of experience of the end users. The cloud paradigm is not the ideal candidate for such latency-sensitive applications owing to the delay between the end user and cloud server. 

This has led to a new trend in computing called Mobile Edge Computing (MEC)~\cite{Satyanarayanan2017, Mao2017}, where the compute capabilities are moved closer to the network edges. It represents an essential building block in the 5G vision of creating large distributed, pervasive, heterogeneous, and multi-domain environments. Harvesting the vast amount of the idle computation power and storage space distributed at the network edges can yield sufficient capacities for performing computation-intensive and latency-critical tasks requested by the end-users. However, it is not feasible to set-up huge resourceful edge clusters along all network edges that mimic the capabilities of the cloud due to the sheer volume of resources that would be required, which would remain underutilized most of the times. Due to the limited resources at the edge nodes and fluctuations in the user requests, an edge cluster may not be capable of meeting the resource and service requirements of all the users it is serving. 

Computation offloading methods have gained a lot of popularity as they provide a simple solution to overcome the problems of edge and mobile computing. Data and computation offloading can potentially reduce the processing delay, improve energy efficiency, and even enhance security for computation-intensive applications. The critical problem in the computation offloading is to determine the amount of computational workload, and choose the MEC server from all available servers. Various aspects of MEC from the point of view of the mobile user have been investigated in the literature. For example, the questions of when to offload to a mobile server, to which mobile server to offload, and how to offload have been studied extensively. See, \cite{Liu2016, Wang2017, Van2018,Chen2018, Wang2019} and references therein. 

However, the design questions at the server side have not been investigated
as extensively. 
When an edge server receives a large number of requests in a short period of time
(for example due to a sporting event),
the edge server can get overloaded, which can lead to service degradation or even node failure. When such service degradation occurs, edge servers are configured to offload requests to other nodes in the cluster in order to avoid the node crash. The crash of an edge node leads to the reduction of the cluster capacity, which is a disaster for the platform operator as well as the end users, who are using the services or the applications. However, performing this migration takes extra time and reduces the resources availability for other services deployed in the cluster. Therefore, it is paramount to design \textit{pro-active} mechanisms that prevent a node from getting overloaded using dynamic offloading policies that can adapt to service request dynamics.

The design of an offloading policy has to take into account the time-varying channel conditions, user mobility, energy supply, computation workload and the computational capabilities of different MEC servers. The problem can be modeled as a Markov Decision Process (MDP) and solved using dynamic programming. However, solving a dynamic program requires the knowledge of the system parameters, which are not typically known and may also vary with time.  In such time-varying environments, the offloading policy must adapt to the environment. Reinforcement Learning (RL) \cite{SuttonBarto2018} is a natural choice to design such adaptive policies as they do not need a model of the environment and can learn the optimal policy based on the observed per-step cost. RL has been successfully applied for designing adaptive offloading policies in edge and fog computing in~\cite{Chen2018RL, huang2019deep, Wang2019RL, li2018deep, chen2018optimized, van2018quality, tang2018migration} to realize one or more objectives such as minimizing latency, minimizing power consumption, association of users and base stations. Although RL has achieved considerable success in the previous work, this success is generally achieved by using deep neural networks to model the policy and the value function. Such deep RL algorithms require considerable computational power and time to train, and are notoriously brittle to the choice of hyper-parameters. They may also not transfer well from simulation to the real-world, and output policies which are difficult to interpret. These features make them impractical to be deployed on the edge nodes to continuously adapt to the changing network conditions.

\begin{figure}[!t]
 \centering
 \includegraphics[width=\linewidth]{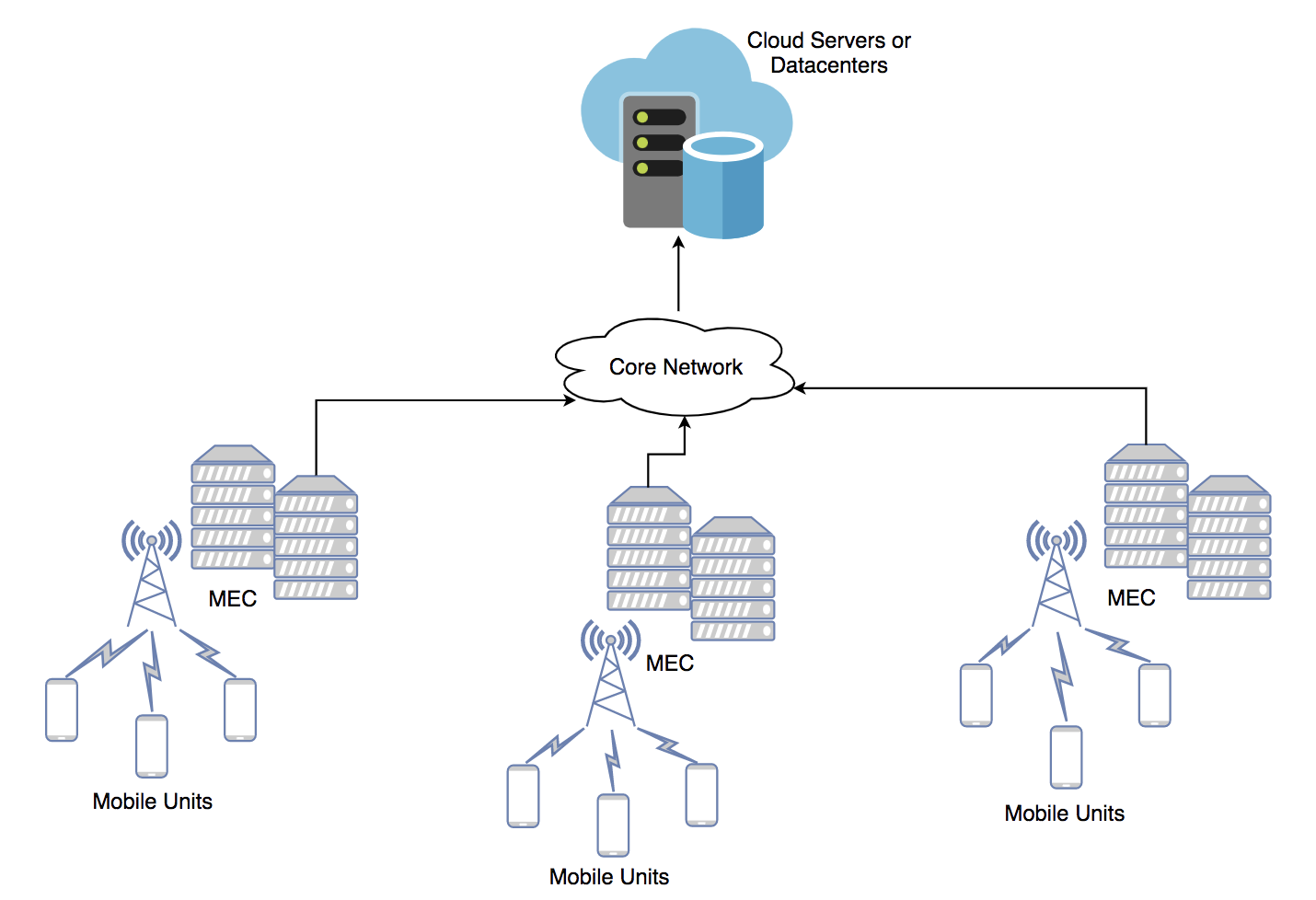}
 \caption{A Mobile Edge Computing (MEC) system.
    User may be mobile and will connect to the closest edge server. The MEC servers are connected to the backend cloud server or datacenters through the core network.}
 \label{fig:model}
\end{figure}

In this work, we study the problem of node-overload protection for a single
edge node. Our main contributions are as follows:

\begin{itemize}
  \item We present a mathematical model for designing an offloading policy for node-overload protection. The model incorporates practical considerations of server holding, processing and offloading costs. In the simplest case, when the request arrival process is time-homogeneous, we model the system as a continuous-time MDP and use the \emph{uniformization technique}~\cite{Jensen1953,Howard1960} to convert the continuous-time MDP to a discrete-time MDP, which can then be solved using standard dynamic programming algorithms~\cite{Puterman1994}.
  \item We show that for time-homogeneous arrival process, the value function and the optimal policy are weakly increasing in the CPU utilization.
  \item We design a node-overload protection scheme that uses a recently proposed low-complexity RL algorithm called Structure-Aware Learning for Multiple Thresholds (SALMUT)~\cite{Roy2019}. The original SALMUT algorithm was designed for the average cost models. We extend the algorithm to the discounted cost setup and prove that SALMUT converges almost surely to a locally optimal policy.
  \item We compare the performance of Deep RL algorithms with SALMUT in a variety of scenarios in a simulated testbed which are motivated by real world deployments. Our simulation experiments show that SALMUT performs close to the state-of-the-art Deep RL algorithms such as PPO~\cite{Schulman2017} and A2C~\cite{Wu2017}, but requires an order of magnitude less time to train and provides optimal policies which are easy to interpret.
  \item We developed a docker testbed where we run actual workloads and compare the performance of SALMUT with the baseline policy. Our results show that SALMUT algorithm outperforms the baseline algorithm.
  \end{itemize}

A preliminary version of this paper appeared in \cite{jitani2021structure}, where the monotonicity results of the optimal policy (Proposition~\ref{prop:value} and \ref{prop:policy}) were stated without proof and the modified SALMUT algorithm was presented with a slightly different derivation. 
However, the convergence behavior of the algorithm (Theorem \ref{th:salmut_convergence}) was not analyzed. A preliminary version of the comparison on SALMUT with state of the art RL algorithms on a computer simulation were included in \cite{jitani2021structure}. However, the detailed behavioral analysis (Sec.~\ref{sec:experiments-1}) and the results for the docker testbed (Sec.~\ref{sec:experiments-2}) are new.

The rest of the paper is organized as follows. We present the system model and
problem formulation in Sec.~\ref{sec:model}. In Sec. \ref{sec:DP}, we present a dynamic
programming decomposition for the case of time-homogeneous statistics of the
arrival process. In Sec.~\ref{sec:RL}, we present the structure aware RL
algorithm (SALMUT) proposed in~\cite{Roy2019} for our model. In Sec.~\ref{sec:experiments-1}, we conduct a detailed experimental study to compare
the performance of SALMUT with other state-of-the-art RL algorithms  using computer simulations. In Sec.~\ref{sec:experiments-2}, we compare the performance of SALMUT with baseline algorithm on the real-testbed. Finally, in Sec.~\ref{sec:conclusion}, we provide the conclusion, limitations of our model, and future directions.
\section{Model and Problem Formulation} \label{sec:model}

\begin{figure}[!t]
 \centering
 \includegraphics[width=\linewidth]{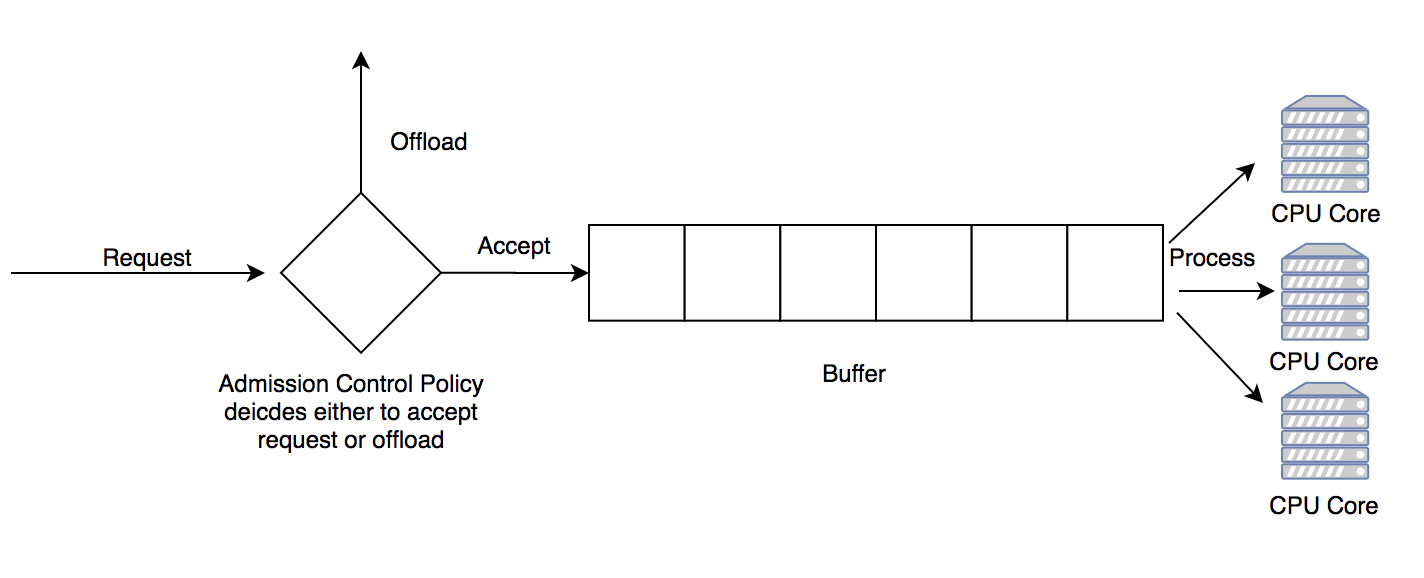}
 \caption{System model of admission control in a single edge server.}
 \label{fig:ad_control}
\end{figure}

\begin{table}[!t]
 \centering
 \caption{List of Symbols Used}
 \label{table:symbols}
 \newcolumntype{R}[1]{>{\RaggedRight\arraybackslash}p{#1}}
 \newcolumntype{L}[1]{>{\RaggedRight\arraybackslash}p{#1}}
 \begin{tabular}{ >{\raggedleft\arraybackslash}p{2cm} R{6.3cm} }
    \toprule
    \textbf{Symbol} & \textbf{Description}\\
    \midrule
    $X_t$ & Queue length at time $t$ \\
    $L_t$ & CPU load of the system at time $t$ \\
    $A_t$ & Offloading action taken by agent at time $t$ \\
    $k$ & Number of cores in the edge node \\
    $R$ & CPU resources required by a request\\
    $P(r)$ & PMF of the CPU resources required\\
    $\mu$ & Processing time of a single core in the edge node\\
    $\lambda$ & Request arrival rate of user\\
    $h$ & Holding cost per unit time \\
    $c(\ell)$ & Running cost per unit time \\
    $p(\ell)$ & Penalty for offloading the packet \\
    $\rho(x, \ell, a)$ & Cost function in the continuous MDP\\
    $\pi$ & Policy of the RL agent \\
    $\alpha$ & Discount factor in continuous MDP \\
    $V^\pi(x,\ell)$ & Performance of the policy $\pi$ \\
    $p(x', \ell' | x, \ell, a)$ & Transition probability function \\
    $\beta$ & Discount factor in discrete MDP \\
    $\bar \rho(x, \ell, a)$ & Cost function in the discrete MDP\\
    $Q(x,\ell,a)$ & Q-value for state $(x, \ell)$ and action $a$ \\
    $\tau$ & Threshold vector \\
    $\pi_\tau$ & Optimal Threshold Policy for SALMUT \\
    $J(\tau)$ & Performance of the SALMULT policy $\pi_\tau$ \\
    $f(\tau(x), \ell)$ & Probability of accepting new request \\
    $T$ & Temperature of the sigmoid function \\
    $\mu(x,\ell)$ & Occupancy measure on the states starting from $(x_0, \ell_0)$ \\
    $b^1_n$ & Fast timescale learning rate\\
    $b^2_n$ & Slow timescale learning rate \\
    $C_{\OFF}$ & Number of requests offloaded by edge node\\
    $C_{\OV}$ & Number of times edge node enters an overloaded state\\
    
    \bottomrule
 \end{tabular}
\end{table}

\subsection{System model}

A simplified MEC system consists of an edge server and several mobile users accessing that server (see Fig.~\ref{fig:model}).
Mobile users independently generate service requests according to a Poisson
process. The rate of requests and the number of users may
also change with time. The edge server takes CPU resources to serve each request from mobile users. 
When a new request arrives, the edge server has the option to serve it or offload it to other healthy edge server in the cluster. The request is buffered in a queue before it is served. 
The mathematical model of the edge server and the mobile
users is presented below.

\subsubsection{Edge server}
Let $X_t \in \{0, 1, \dots, \mathsf{X}\}$ denote the number of service
requests buffered in the queue, where $\mathsf{X}$ denotes the size of the
buffer. Let $L_t \in \{0, 1, \dots, \mathsf{L}\}$ denote the CPU load at the
server where $\mathsf{L}$ is the capacity of the CPU. We assume that the CPU
has $k$ cores.

We assume that the requests arrive according to a (potentially time-varying)
Poisson process with rate~$\lambda$. If a new request arrives when the buffer is full, the request
is offloaded to another server. If a new request arrives when the buffer is
not full, the server has the option to either accept or offload the request. 

The server can process up to a maximum of $k$ requests from the head of the
queue. Processing each request requires CPU resources for the duration in
which the request is being served. The required CPU resources is a random
variable $R \in \{1, \dots, \mathsf{R}\}$ with probability mass function
$P$. The realization of $R$ is not revealed until the server starts working on
the request. The duration of service is exponentially distributed random
variable with rate~$\mu$.

Let $\mathcal{A} = \{0, 1\}$ denote the action set. Here $A_t = \REJECT$ means that
the server decides to offload the request while $A_t = \ACCEPT$ means that the
server accepts the request. 

\subsubsection{Traffic model for mobile users} We consider multiple models for
traffic.

\begin{itemize}
  \item \textbf{Scenario 1:} All users generate requests according to the same
    rate $\lambda$ and the rate does not change over time. Thus, the rate at which
    requests arrive is $\lambda N$.
 \item \textbf{Scenario 2:} In this scenario, we assume that all users
    generate requests according to rate $\lambda_{M_t}$, where $M_t \in \{1, \dots, \mathsf{M}\}$ is a global
    state which changes over time. Thus, the rate at which
    requests arrive in state~$m$, where $m \in M_t$, is $\lambda_m N$. 
 \item \textbf{Scenario 3:} Each user~$n$ has a state $M^n_t \in \{1, \dots,
    \mathsf{M}\}$. When the user $n$ is in state~$m$, it generates requests
    according to rate $\lambda_m$. The state $M^n_t$ changes over time.
    Thus, the rate at which requests arrive at the
    server is $\sum_{n=1}^N \lambda_{M^n_t}$. 
  \item \textbf{Time-varying user set:} In each of the scenarios above, we can
    consider the case when the number of users is not fixed and changes over
    time. We call them Scenario 4, 5, and 6 respectively.
\end{itemize}

\subsubsection{Cost and the optimization framework}

The system incurs three types of a cost:
\begin{itemize}
  \item a holding cost of $h$ per unit time when a request is buffered in the
    queue but is not being served. 
  \item a running cost of $c(\ell)$ per unit time for running the CPU at a
    load of $\ell$. 
  \item a penalty of $p(\ell)$ for offloading a packet at CPU load $\ell$. 
\end{itemize}
We combine all these costs in a cost function
\begin{equation}\label{eq:cost}
  \rho(x,\ell,a) = h[x-k]^{+} + c(\ell) + p(\ell) \IND\{a = \REJECT\},
\end{equation}
where $a$ denotes the action, $[x]^{+}$ is a short-hand for $\max\{x,0\}$ and $\IND\{ \cdot\}$ is the
indicator function.
Note that to simplify the analysis, we assume that the server always
serves $\min\{ X_t, k\}$ requests. It is also assumed that $c(\ell)$ and $c(\ell) +
p(\ell)$ are increasing in $\ell$. 

Whenever a new request
arrives, the server uses a memoryless policy $\pi \colon \{0, 1, \dots,
\mathsf{X} \} \times \{0, 1, \dots, \mathsf{L}\} \to \{0, 1\}$ to choose an
action
\[
  A_t = \pi_t(X_t, L_t).
\]

The performance of a policy $\pi$ starting from initial state $(x, \ell)$ is
given by 
%
\begin{equation}\label{eq:performance}
  V^\pi(x,\ell) = \EXP \biggl[
    \int_{0}^\infty e^{-\alpha t} \rho(X_t, L_t, A_t) dt 
    \biggm| X_0 = x, L_0 = \ell
  \biggr],
\end{equation}
where $\alpha > 0$ is the discount rate and the expectation is with respect to the
arrival process, CPU utilization, and service completions. 

The objective is to minimize the performance~\eqref{eq:performance} for the
different traffic scenarios listed above. We are particularly interested in
the scenarios where the arrival rate and potentially other components of the
model such as the resource distribution are not known to the system designer
and change during the operation of the system.

\subsection{Solution framework}

When the model parameters $(\lambda, N, \mu, P, k)$ are
known and time-homogeneous, the optimal policy~$\pi$ can be computed using dynamic programming.
However, in a real system, these parameters may not be known, so we are
interested in developing a RL algorithm which can learn
the optimal policy based on the observed per-step cost. 

In principle, when the model parameters are known, Scenarios~2 and~3 can also
be solved using dynamic programming. However, the state of such dynamic
programs will include the state~$M_t$ of the system (for Scenario~2) or the states
$(M^n_t)_{n =1}^N$ of all users (for Scenario~3). Typically, these states
change at a slow time-scale. So, we will consider reinforcement learning
algorithms which do not explicitly keep track of the states of the user and
verify that the algorithm can adapt quickly whenever the arrival rates change.

\section{Dynamic programming to identify optimal admission control policy} \label{sec:DP}

When the arrival process is time-homogeneous, the process $\{X_t, L_t\}_{t \ge 0}$ is a finite-state
continuous-time MDP controlled through $\{A_t\}_{t
\ge 0}$. To specify the controlled transition
probability of this MDP, we consider the following two cases. 

First, if there is a new arrival at time~$t$, then 
\begin{align}
  \hskip 2em & \hskip -2em
  \PR(X_{t} = x', L_t = \ell' \mid X_{t^{-}} = x, L_{t^{-}} = \ell, A_{t} = a)
  \notag \\
  &= \begin{cases}
    P(\ell' - \ell), & \text{if $x' = x + 1$ and $a = \ACCEPT$} \\
    1, & \text{if $x' = x$, $\ell' = \ell$, and $a = \REJECT$} \\
    0, & \text{otherwise}.
  \end{cases} \label{eq:arrival}
\end{align}
We denote this transition function by $q_{+}(x',\ell' | x, \ell, a)$. 
Note that the first term $P(\ell'-\ell)$ denotes the probability that the
accepted request required $(\ell' - \ell)$ CPU resources. 

Second, if there is a departure at time~$t$,
\begin{align}
  \hskip 2em & \hskip -2em
  \PR(X_{t} = x', L_t = \ell' \mid X_{t^{-}} = x, L_{t^{-}} = \ell)
  \notag \\
  &= \begin{cases}
    P(\ell - \ell'), & \text{if $x' = [x - 1]^{+}$} \\
    0, & \text{otherwise}.
  \end{cases} \label{eq:departure}
\end{align}
We denote this transition function by $q_{-}(x',\ell' | x, \ell)$. 
Note that there is no decision to be taken at the completion of a request, so
the above transition does not depend on the action.
In
general, the reduction in CPU utilization will correspond to the resources
released after the client requests are served. However, keeping track
of those resources would mean that we would need to expand the state and
include $(R_1, \dots, R_k)$ as part of the state, where $R_i$ denotes the
resources required by the request which is being processed by CPU~$i$. In
order to avoid such an increase in state dimension, we assume that when a
request is completed, CPU utilization reduces by amount $\ell - \ell'$ with
probability $P(\ell - \ell')$. 

We combine~\eqref{eq:arrival} and~\eqref{eq:departure} into a single
controlled transition probability function from state $(x,\ell)$ to state
$(x',\ell')$ given by
\begin{align}
  p(x', \ell' \mid x, \ell, a) &=
  \frac{\lambda}{\lambda + \min\{x,k\}\mu} q_{+}(x',\ell' \mid x, \ell, a)
  \notag\\
  & \quad + 
  \frac{\min\{x,k\}\mu}{\lambda + \min\{x,k\}\mu} q_{-}(x',\ell' \mid x, \ell).
\end{align}

Let $\nu = \lambda + k\mu$ denote the uniform upper bound on the transition
rate at the states. Then, using the \emph{uniformization
technique}~\cite{Jensen1953,Howard1960}, we can
convert the above continuous time discounted cost MDP into a discrete time
discounted cost MDP with discount factor \( \beta = \nu/(\alpha + \nu) \),
transition probability matrix $p(x',\ell' | x,\ell, a)$ and per-step cost
\[
  \bar \rho(x, \ell, a) = \frac{1}{\alpha + \nu} \rho(x, \ell, a).
\]

Therefore, we have the following.
\begin{theorem}\label{th:DP}
  Consider the following dynamic program
  \begin{equation}\label{eq:DP}
    V(x, \ell) = \min\{ Q(x,\ell, \ACCEPT), Q(x,\ell,\REJECT) \}
  \end{equation}
  where
  \begin{align*}
    Q(x,&\ell,\ACCEPT) = \frac{1}{\alpha + \nu} \bigl[ h[x-k]^{+} + c(\ell) \bigr]
    \notag \\
    &  
    + \beta \bigg[
      \frac{\lambda}{\lambda + \min\{x,k\}\mu} \sum_{r = 1}^{\mathsf{R}}P(r)
      V([x+1]_{\mathsf{X}}, [\ell + r]_{\mathsf{L}})
    \notag \\
    & \qquad + 
      \frac{\min\{x,k\}\mu }{\lambda + \min\{x,k\}\mu} \sum_{r = 1}^{\mathsf{R}}P(r)
    V([x-1]^{+}, [\ell - r]^{+}) \biggr]
    \shortintertext{and}
    Q(x,&\ell,\REJECT) = \frac{1}{\alpha + \nu} \bigl[ h[x-k]^{+} + c(\ell) + p(\ell)  \bigr]
    \notag \\
    &  
    + \beta 
      \frac{\min\{x,k\}\mu }{\lambda + \min\{x,k\}\mu} \sum_{r = 1}^{\mathsf{R}}P(r)
      V([x-1]^{+}, [\ell - r]^{+})
  \end{align*}
  where $[x]_{\mathsf{B}}$ denotes $\min\{x,\mathsf{B}\}$. 

  Let $\pi(x,\ell) \in \mathcal{A}$ denote the argmin the right hand side
  of~\eqref{eq:DP}. Then, the time-homogeneous policy $\pi(x,\ell)$ is optimal
  for the original continuous-time optimization problem.
\end{theorem}
\begin{proof}
  The equivalence between the continuous and discrete time MDPs follows from
  the uniformization technique~\cite{Jensen1953,Howard1960}. The optimality of
  the time-homogeneous policy $\pi$ follows from the standard results for
  MDPs~\cite{Puterman1994}.
\end{proof}

Thus, for all practical purposes, the decision maker has to solve a
discrete-time MDP, where he has to take decisions at the instances when a new
request arrives. In the sequel, we will ignore the $1/(\alpha + \nu)$ term in
front of the per-step cost and assume that it has been absorbed in the
constant $h$, and the functions $c(\cdot)$, $p(\cdot)$. 

When the system parameters are known, the above dynamic program can be solved
using standard techniques such as value iteration, policy iteration, or linear
programming. However, in practice, the system parameters may slowly change
over time. Therefore, instead of pursuing a planning solution, we consider
reinforcement learning solutions which can adapt to time-varying environments.

\section{Structure-aware reinforcement learning} \label{sec:RL}

Although, in principle, the optimal admission control problem formulated above can be
solved using deep RL algorithms, such algorithms require significant
computational resources to train, are brittle to the choice of
hyperparameters, and generate policies which are difficult to interpret. For
the aforementioned reasons, we investigate an alternate class of RL algorithms which circumvents
these limitations. 

\subsection{Structure of the optimal policy}

We first establish basic monotonicity properties of the value function and the
optimal policy.

\begin{proposition}\label{prop:value}
    For a fixed queue length $x$, the value function is weakly
      increasing in the CPU utilization $\ell$.
\end{proposition}

\begin{proof}
The proof is present in Appendix \ref{appendix:A}.
\end{proof}

\begin{proposition}\label{prop:policy}
    For a fixed queue length $x$, if it is optimal to reject a request
      at CPU utilization $\ell$, then it is optimal to reject a request at all
      CPU utilizations $\ell' > \ell$.
\end{proposition}

\begin{proof}
The proof is present in Appendix \ref{appendix:B}.
\end{proof}


\subsection{The SALMUT algorithm}

Proposition~\ref{prop:policy} shows that the optimal policy can be represented
by a threshold vector $\tau = (\tau(x))_{x=0}^{\mathsf{X}}$, where $\tau(x) \in
\{0, \dots, \mathsf{L}\}$ is the smallest value of the CPU utilization such
that it is optimal to accept the packet for CPU utilization less than or equal to
$\tau(x)$ and reject it for utilization greater than $\tau(x)$.

The SALMUT algorithm was proposed in~\cite{Roy2019} to exploit a similar
structure in admission control for multi-class queues. It was originally proposed for the
average cost setting. We present a generalization to the discounted-time setting. 

We use $\pi_\tau$ to denote a threshold-based policy with the
parameters $(\tau(x))_{x = 0}^{\mathsf{X}}$ taking values in $\{0, \dots,
\mathsf{L}\}^{\mathsf{X} + 1}$. The key idea behind SALMUT is that, instead of
deterministic threshold-based policies, we consider a random policy parameterized with parameters taking value in the compact set $[0,
\mathsf{L}]^{\mathsf{X} + 1}$. Then, for any state $(x,\ell)$, the randomized
policy $\pi_\tau$ chooses action $a = \ACCEPT$ with probability $f(\tau(x),\ell)$
and chooses action $a = \REJECT$ with probability $1 - f(\tau(x), \ell)$, where
$f(\tau(x), \ell)$ is any continuous decreasing function w.r.t $\ell$, which is
differentiable in its first argument, e.g., the sigmoid function
\begin{equation} \label{eq:randomized}
  f(\tau(x), \ell) = \frac{ \exp( (\tau(x) - \ell)/T ) }
  {  1 +\exp( (\tau(x) - \ell)/T ) },
\end{equation}
where $T > 0$ is a hyper-parameter (often called ``temperature''). 



Fix an initial state $(x_0, \ell_0)$ and let $J(\tau)$ denote the performance
of policy $\pi_\tau$. Furthermore, let $p_\tau(x',\ell'|x,\ell)$ denote the transition function under policy $\pi_\tau$, i.e.

\begin{multline}
    \label{eq:transition_prob}
    p_\tau(x',\ell'|x,\ell) = f(\tau(x), \ell) p(x',\ell'|x,\ell,0) \\ + (1 - f(\tau(x), \ell)) p(x',\ell'|x,\ell,1)
\end{multline}

Similarly, let $\bar \rho_\tau(x,\ell)$ denote the expected per-step reward under policy $\pi_\tau$, i.e.
\begin{multline}
    \label{eq:cost_fn}
    \bar \rho_\tau(x,\ell) = f(\tau(x), \ell) \bar \rho (x,\ell,0)  + (1 - f(\tau(x), \ell)) \bar \rho(x,\ell,1).
\end{multline}
Let $\nabla$ denote the gradient with respect to $\tau$.

From Performance Derivative formula \cite[Eq. 2.44]{cao2007stochastic}, we know that
\begin{multline}
  \label{eq:grad_J_tau}
  \GRAD J(\tau) =
  \frac{1}{1 - \beta} \sum_{x = 0}^{\mathsf{X}} \sum_{\ell=0}^{\mathsf{L}} \mu(x,\ell) \bigl[\nabla \bar \rho_\tau(x,\ell) \\
  + \beta \sum_{x' = 0}^{\mathsf{X}} \sum_{\ell'=0}^{\mathsf{L}} \nabla p_\tau(x',\ell'|x,\ell) V_\tau(x',\ell')]
\end{multline}
where $\mu(x,\ell)$ is the occupancy measure on the states starting from the initial state $(x_0, \ell_0)$.

From \eqref{eq:transition_prob}, we get that 
\begin{multline}
    \nabla p_\tau(x',\ell'|x,\ell) = (p(x',\ell'|x,\ell,0) - p(x',\ell'|x,\ell,1)) \\ \nabla f(\tau(x), \ell).
\end{multline}
Similarly, from \eqref{eq:cost_fn}, we get that 
\begin{equation}
    \nabla \bar \rho_\tau(x,\ell) = (\bar \rho (x,\ell,0) - \bar \rho (x,\ell,1)) \\ \nabla f(\tau(x), \ell).
\end{equation}
Substituting \eqref{eq:transition_prob} \& \eqref{eq:cost_fn} in \eqref{eq:grad_J_tau} and simplifying, we get
\begin{multline}
  \label{eq:grad_J_tau_simp}
  \GRAD J(\tau) =
  \frac{1}{1 - \beta} \sum_{x = 0}^{\mathsf{X}} \sum_{\ell=0}^{\mathsf{L}} \mu(x,\ell) [\Delta Q(x,\ell)] \nabla f(\tau(x), \ell),
\end{multline}
where $\Delta Q(x,\ell) = Q(x,\ell,0) - Q(x,\ell,1)$.




Therefore, when $(x,\ell)$ is sampled from the stationary distribution $\mu$, an unbiased estimator of $\GRAD J(\tau)$ is proportional to $\Delta Q(x,\ell) \nabla f(\tau(x), \ell)$.


\begin{figure}[!t]
 \centering
 \includegraphics[width=\linewidth]{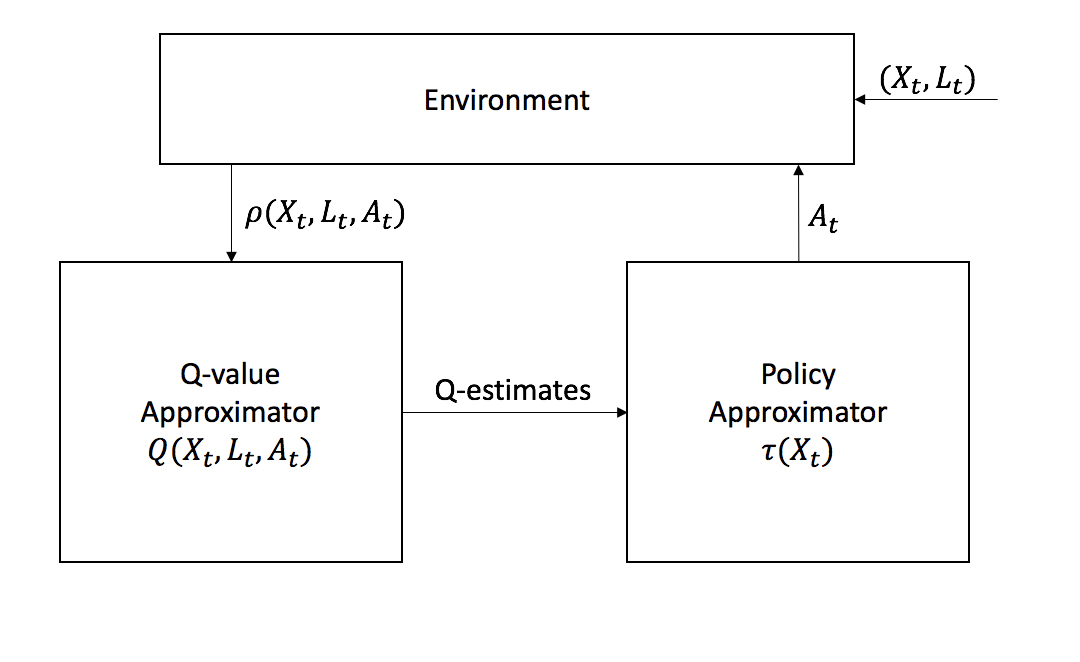}
 \caption{Illustration of the two-timescale SALMUT algorithm.}
 \label{fig:two-timescale}
\end{figure}

Thus, we can use the standard two time-scale Actor-Critic
algorithm~\cite{SuttonBarto2018} to simultaneously learn the policy parameters
$\tau$ and the action-value function $Q$ as follows. We start with an initial
guess $Q_0$ and $\tau_0$ for the action-value function and the optimal policy parameters.
Then, we update the action-value function using temporal difference learning:
\begin{multline}
  Q_{n+1}(x,\ell,a) = Q_n(x,\ell,a) + b^1_n\bigl[ \bar \rho(x,\ell,a) 
    \\ +
  \beta \min_{a' \in A} Q_n(x',\ell',a') - Q_n(x,\ell,a) \bigr],
 \label{eq:q_val_update}
\end{multline}
and update the policy parameters using stochastic gradient descent while
using the unbiased estimator of $\GRAD J(\tau)$:

\begin{equation}
  \tau_{n+1}(x) = \PROJ\bigl[
    \tau_n(x) + b^{2}_n \GRAD f(\tau(x), \ell) \Delta Q(x,\ell)],
 \label{eq:thres_update}
\end{equation}
where $\PROJ$ is a projection operator which clips the values to the interval
$[0, \mathsf{L}]$ and $\{b^1_n\}_{n \ge 0}$ and $\{b^2_n\}_{n \ge 0}$ are
learning rates which satisfy the standard conditions on two time-scale
learning: $\sum_{n} b^k_n =\infty$, $\sum_n (b^k_n)^2 < \infty$, $k \in \{1,
2\}$, and $\lim_{n \to \infty} b^2_n/b^1_n = 0$. 

\begin{algorithm}
\SetAlgoLined
\KwResult{$\tau$}
 Initialize action-value function $\forall x, \forall \ell$, $Q(x, \ell, a) \leftarrow 0$ \\
 Initialize threshold vector  $\forall x$, $\tau(x) \leftarrow
 \text{rand}(0,\mathsf{L})$  \\
 Initialize start state $(x, \ell) \leftarrow (x_0, \ell_0)$  \\
 \While{\textsc{true}}{
   \If{\textsc{event} == \textsc{arrival}}{
   Choose action $a$ according to Eq.~\eqref{eq:randomized} \\
   Observe next state $(x',\ell')$ \\
   Update $Q(x, \ell, a)$ according to Eq.~\eqref{eq:q_val_update} \\
   Update threshold $\tau$ using Eq.~\eqref{eq:thres_update} \\
   $(x, \ell) \xleftarrow{} (x', \ell')$ 
   }
 }
 \caption{Two time-scale SALMUT algorithm}
 \label{alg:SALMUT}
\end{algorithm}

The complete algorithm is presented in Algorithm~\ref{alg:SALMUT} and illustrated in Fig.~\ref{fig:two-timescale}.
\begin{theorem}\label{th:salmut_convergence}
The two time-scale SALMUT algorithm described above converges almost surely and
$\lim_{n \rightarrow \infty}\GRAD J(\tau_n) = 0$.
\end{theorem}
\begin{proof}
The proof is present in Appendix \ref{appendix:C}.
\end{proof}

\begin{remark}\label{remark}
The idea of replacing the "hard" threshold $\tau(x) \in \{0, \dots, \mathsf{L}\}^{\mathsf{X+1}}$ with a "soft" threshold $\tau(x) \in [0, \mathsf{L}]^{\mathsf{X+1}}$ is same as that of the SALMUT algorithm \cite{Roy2019}. However, our simplification of the performance derivative \eqref{eq:grad_J_tau} given by \eqref{eq:grad_J_tau_simp} is conceptually different from the simplification presented in \cite{Roy2019}. The simplification in \cite{Roy2019} is based on viewing
$\sum_{x' = 0}^{\mathsf{X}} \sum_{\ell'=0}^{\mathsf{L}} \nabla p_\tau(x',\ell'|x,\ell) V_\tau(x',\ell')$ term in \eqref{eq:grad_J_tau} as
\[
2 \mathds{E} \bigl[(-1)^{\delta} \nabla f(\tau(x), \ell) V_\tau(\hat x, \hat \ell)]
\]
where $\delta \sim \text{Unif} \{0,1\}$ is an independent binary random variable and $(\hat x, \hat \ell) \sim \delta p(\cdot|x,\ell,0) + (1 - \delta) p(\cdot|x,\ell, 1)$. In contrast, our simplification is based on a different algebric simplification that directly simplifies \eqref{eq:grad_J_tau} without requiring any additional sampling. 
\end{remark}

\section{Numerical experiments - Computer Simulations}\label{sec:experiments-1}

In this section, we present detailed numerical experiments to evaluate the
proposed reinforcement learning algorithm on various scenarios described in Sec.~\ref{sec:model}-A.

We consider an edge server with buffer size
$\mathsf{X} = 20$, CPU capacity $\mathsf{L} = 20$, $k = 2$ cores, service-rate $\mu = 3.0$
for each core, holding cost $h = 0.12$. The CPU capacity is discretized into
$20$ states for utilization $0-100 \%$, with $\ell = 0$ corresponding to a
state with CPU load $\ell \in [0\%-5\%)$, and so on. 

The CPU running cost is modelled such that it incurs a positive reinforcement for being in the optimal CPU range, and a high cost for an overloaded system.
\[
  c(\ell) = \begin{cases}
    0  & \text{for $\ell \le 5$} \\
    -0.2 & \text{for $6 \le \ell \le 17$} \\
    10 & \text{for $\ell \ge 18$}
  \end{cases}
\]

The offload penalty is modelled such that it incurs a fixed cost for offloading to enable the offloading behavior only when the system is loaded and a very high cost when load is system is idle to discourage offloading in such scenarios.
\[
  p(\ell) = \begin{cases}
    1  & \text{for $\ell \ge 3$} \\
    10 & \text{for $\ell \le 3$}
  \end{cases}
\]
The  probability
mass function of resources requested per request is as follows
\[
  P(r) = \begin{cases}
    0.6 & \text{if $r = 1$} \\
    0.4 & \text{if $r = 2$} 
  \end{cases}.
\]

Rather than simulating the system in continuous-time, we simulate the
equivalent discrete-time MDP by generating the next event (arrival or departure) using a Bernoulli distribution with probabilities and costs described in Sec.~\ref{sec:DP}. We assume that the parameter $1/(\alpha + \nu)$ in~\eqref{eq:DP} has been absorbed in the cost function. We assume that the discrete time discount factor $\beta =
\alpha/(\alpha + \nu)$ equals $0.95$.

\subsection{Simulation scenarios}
\begin{figure}[!htb]
    \centering
    \begin{subfigure}[t]{0.48\linewidth}
      \centering
      \includegraphics[width=\textwidth]{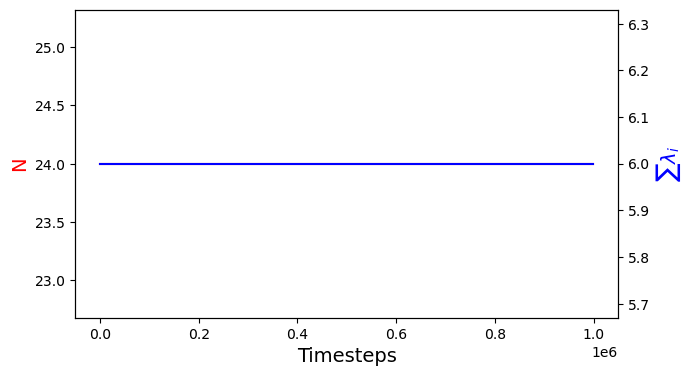}
      \caption{Scenario 1}
      \label{fig:n_v_lambda_scenario1}
    \end{subfigure}%
    \hfill
    \begin{subfigure}[t]{0.48\linewidth}
      \centering
      \includegraphics[width=\textwidth]{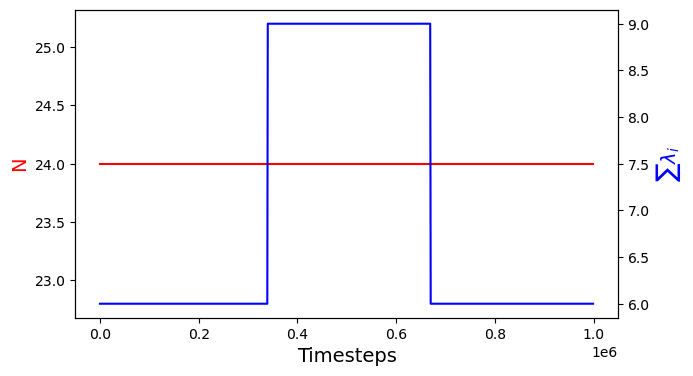}
      \caption{Scenario 2}
      \label{fig:n_v_lambda_scenario2}
    \end{subfigure}%
    \hfill
    \begin{subfigure}[t]{0.48\linewidth}
      \centering
      \includegraphics[width=\textwidth]{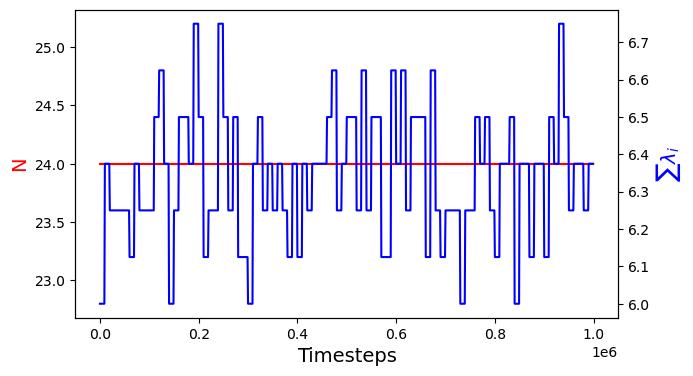}
      \caption{Scenario 3}
      \label{fig:n_v_lambda_scenario3}
    \end{subfigure}%
    \hfill
    \begin{subfigure}[t]{0.48\linewidth}
      \centering
      \includegraphics[width=\textwidth]{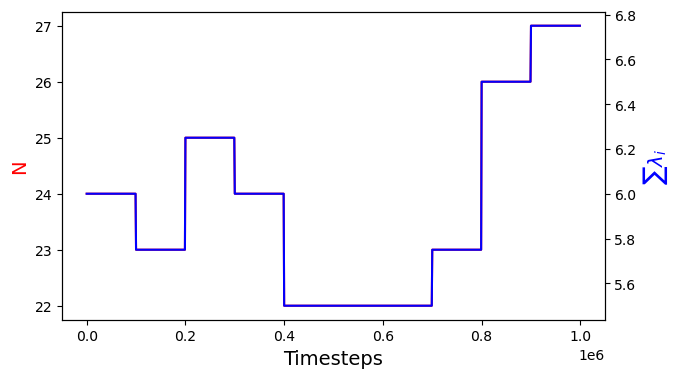}
      \caption{Scenario 4}
      \label{fig:n_v_lambda_scenario4}
    \end{subfigure}%

    \begin{subfigure}[t]{0.48\linewidth}
      \centering
      \includegraphics[width=\textwidth]{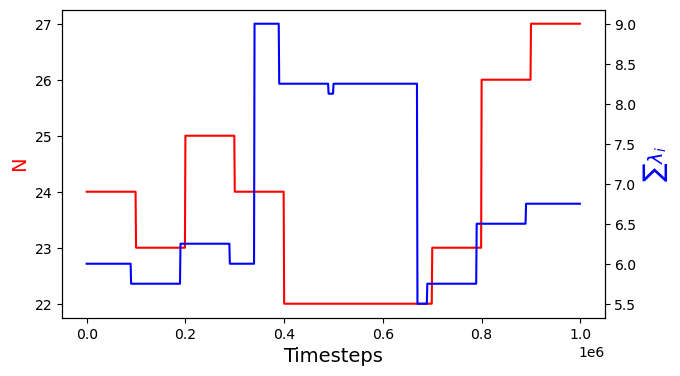}
      \caption{Scenario 5}
      \label{fig:n_v_lambda_scenario5}
    \end{subfigure}%
    \hfill
    \begin{subfigure}[t]{0.48\linewidth}
      \centering
      \includegraphics[width=\textwidth]{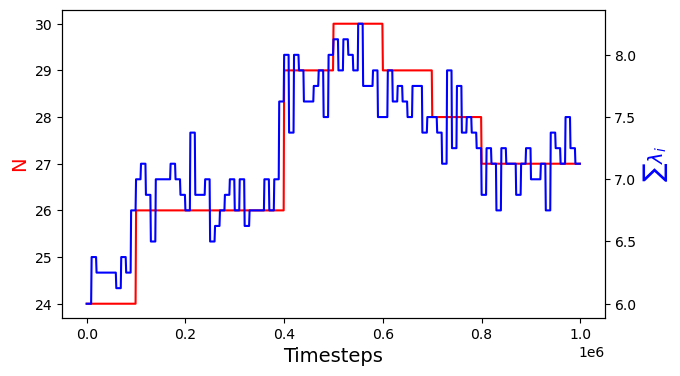}
      \caption{Scenario 6}
      \label{fig:n_v_lambda_scenario6}
    \end{subfigure}%
    \caption{The evolution of $\lambda$ and $N$ for the different scenarios
    that we described. In scenarios 1 and 4, $\lambda$ and $N$ overlap in the plots.}
    \label{fig:evolution}
\end{figure}

We consider a number of traffic scenarios which are increasing in complexity and
closeness to the real-world setting. Each scenario runs for a horizon of $T =10^{6}$. The scenarios capture variation in the transmission rate and the number of users over time, their realization can be seen in Fig.~\ref{fig:evolution}.

The evolution of the arrival rate $\lambda$ and the number of users
$N$ for the more dynamic environments is shown in Fig.~\ref{fig:evolution}.

\paragraph{Scenario 1} This scenario tests how the learning algorithms perform
in the time-homogeneous setting. We consider a system with $N=24$ users with
arrival rate $\lambda_i = 0.25$. Thus, the overall arrival rate $\lambda = N
\lambda_i = 6$.

\paragraph{Scenario 2} This scenario tests how the learning algorithms adapt
to occasional but significant changes to arrival rates. We consider a system
with $N = 24$ users, where each user generates requests at rate
$\lambda_{\text{low}} = 0.25$ for the interval $(0, 3.33 \times 10^{5}]$, then
generates requests at rate $\lambda_{\text{high}} = 0.375$ for the interval
$(3.34 \times 10^{5}, 6.66 \times 10^{5}]$, and then generates requests at
rate $\lambda_{\text{low}}$ again for the interval $(6.67 \times 10^5, 10^{6}]$.

\paragraph{Scenario 3} This scenario tests how the learning algorithms adapt
to frequent but small changes to the arrival rates. We consider a system with
$N = 24$ users, where each user generates requests according to rate $\lambda
\in \{ \lambda_{\text{low}}, \lambda_{\text{high}}\}$ where
$\lambda_{\text{low}} = 0.25$ and $\lambda_{\text{high}} = 0.375$. We assume
that each user starts with a rate $\lambda_{\text{low}}$ or
$\lambda_{\text{high}}$ with equal probability. At time intervals $m \times
10^4$, each user toggles its transmission rate with probability $p = 0.1$. 

\paragraph{Scenario 4} This scenario tests how the learning algorithm adapts
to change in the number of users. In particular, we consider a setting where
the system starts with $N_1 = 24$ user. At every $10^5$ time steps, a user may leave the network, stay in the network or add another mobile device to the network with
probabilities $0.05$, $0.9$, and $0.05$, respectively. Each new user generates
requests at rate~$\lambda$.

\paragraph{Scenario 5} This scenario tests how the learning algorithm adapts
to large but occasional change in the arrival rates and small changes in the
number of users. In particular, we consider the setup of Scenario~2, where the
number of users change as in Scenario~4.

\paragraph{Scenario 6} This scenario tests how the learning algorithm adapts
to small but frequent change in the arrival rates and small changes in the
number of users. In particular, we consider the setup of Scenario~3, where the
number of users change as in Scenario~4.

\subsection{The RL algorithms}
For each scenarios, we compare the performance of the following policies
\begin{enumerate}
    \item Dynamic Programming (DP), which computes the optimal policy using Theorem~\ref{th:DP}.
    \item SALMUT, as described in Sec.~\ref{sec:RL}-B.
    \item Q-Learning, using \eqref{eq:q_val_update}.
    \item PPO \cite{Schulman2017}, which is a family of trust region policy gradient method and optimizes a surrogate objective function using stochastic gradient ascent.
    \item A2C \cite{Wu2017}, which is a
    two time-timescale learning algorithms where the critic estimates the value function and actor updates the policy distribution in the direction suggested by the critic.
    \item Baseline, which is a fixed-threshold based policy, where the node accepts requests when $\ell < 18$ (non-overloaded state) and offloads requests otherwise. Such static policies are currently deployed in many real-world systems.
\end{enumerate}

For SALMUT, we use ADAM \cite{kingma2014adam} optimizer with initial learning rates $(b^1 = 0.03, b^2 = 0.002)$. For Q-learning, we use Stochastic Gradient Descent with $b^1 = 0.01$. We used the stable-baselines \cite{stable-baselines3} implementation of PPO and A2C with learning rates $0.0003$ and $0.001$ respectively. 

\subsection{Results}

\begin{figure*}[!t]
    \centering
    \begin{subfigure}[t]{0.33\linewidth}
      \centering
      \includegraphics[width=\textwidth]{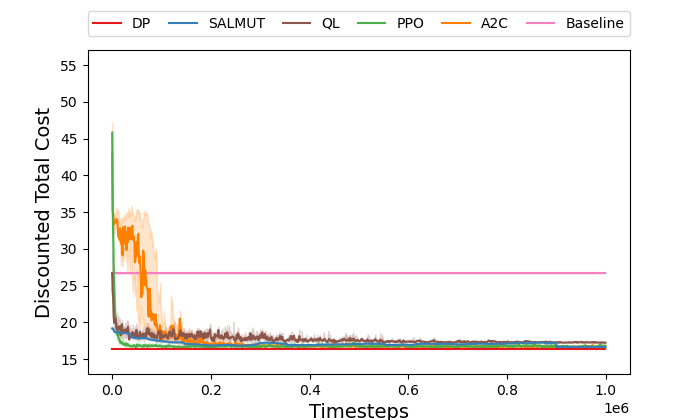}
      \caption{Scenario 1}
      \label{fig:scenario1}
    \end{subfigure}%
    \hfill
    \begin{subfigure}[t]{0.33\linewidth}
      \centering
      \includegraphics[width=\textwidth]{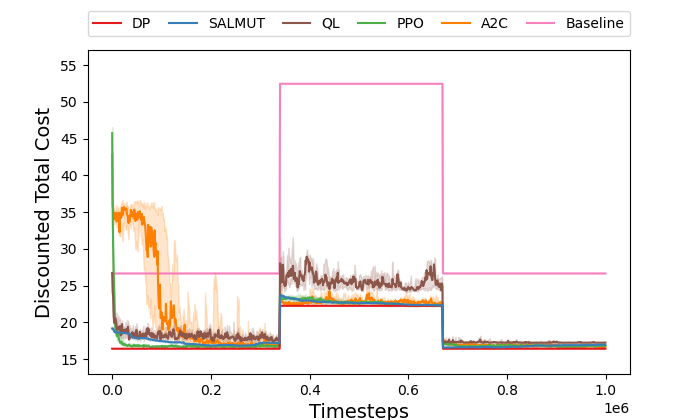}
      \caption{Scenario 2}
      \label{fig:scenario2}
    \end{subfigure}%
    \hfill
    \begin{subfigure}[t]{0.33\linewidth}
      \centering
      \includegraphics[width=\textwidth]{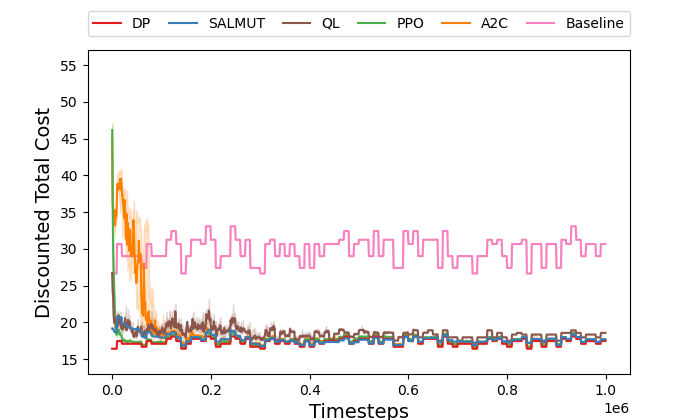}
      \caption{Scenario 3}
      \label{fig:scenario3}
    \end{subfigure}%
 
    \begin{subfigure}[t]{0.33\linewidth}
      \centering
      \includegraphics[width=\textwidth]{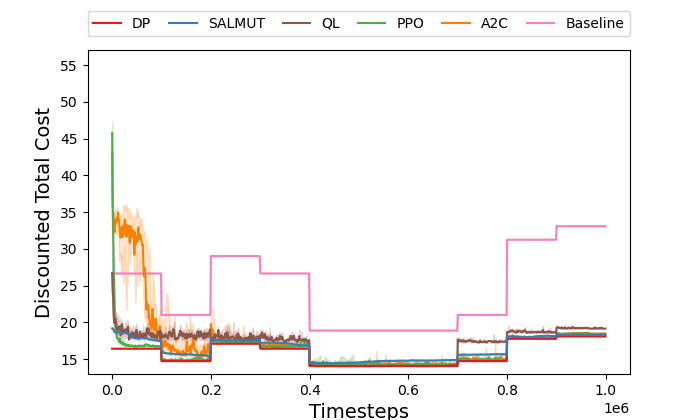}
      \caption{Scenario 4}
      \label{fig:scenario4}
    \end{subfigure}%
    \hfill
    \begin{subfigure}[t]{0.33\linewidth}
      \centering
      \includegraphics[width=\textwidth]{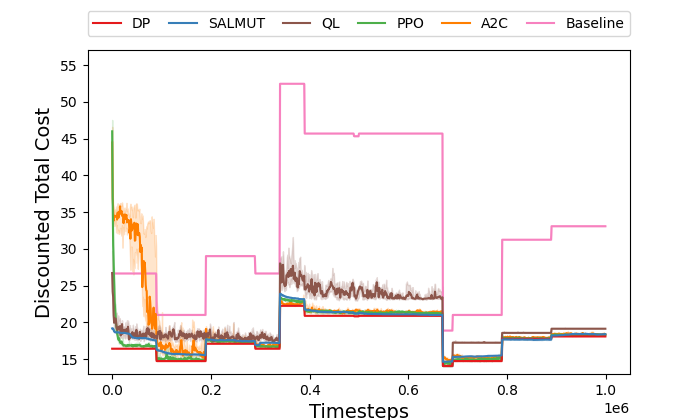}
      \caption{Scenario 5}
      \label{fig:scenario5}
    \end{subfigure}%
    \hfill
    \begin{subfigure}[t]{0.33\linewidth}
      \centering
      \includegraphics[width=\textwidth]{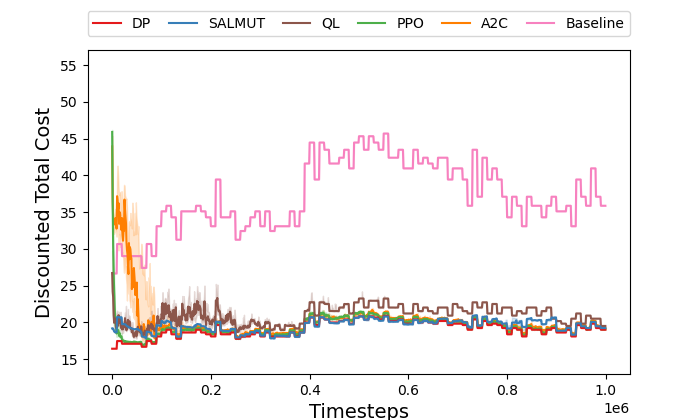}
      \caption{Scenario 6}
      \label{fig:scenario6}
    \end{subfigure}%
    \caption{Performance of RL algorithms for different scenarios.} 
    \label{fig:results}
\end{figure*}

For each of the algorithm described above, we train SALMUT, Q-learning, PPO, and A2C for $10^6$ steps. The performance of each algorithm is
evaluated every $10^3$ steps using independent rollouts of length $H = 1000$ for $100$ different random seeds. The experiment is repeated for the $10$ sample paths and the median performance with an uncertainty band from the first to the third quartile are plotted in Fig.~\ref{fig:results}. 

For Scenario~1, all RL algorithms (SALMUT, Q-learning, PPO, A2C) converge to a
close-to-optimal policy and remain stable after convergence. Since all policies converge quickly, SALMUT, PPO, and A2C
are also able to adapt quickly in Scenarios~2--6 and keep track of the
time-varying arrival rates and number of users. There are small differences in
the performance of the RL algorithms, but these are minor. Note that, in
contrast, Q-learning policy does not perform well when the dynamics of the requests changes drastically, whereas the baseline policy performs poorly when the server is overloaded. 

\begin{figure}[!t]
    \centering
    \begin{subfigure}[t]{0.5\linewidth}
      \centering
      \includegraphics[width=\textwidth]{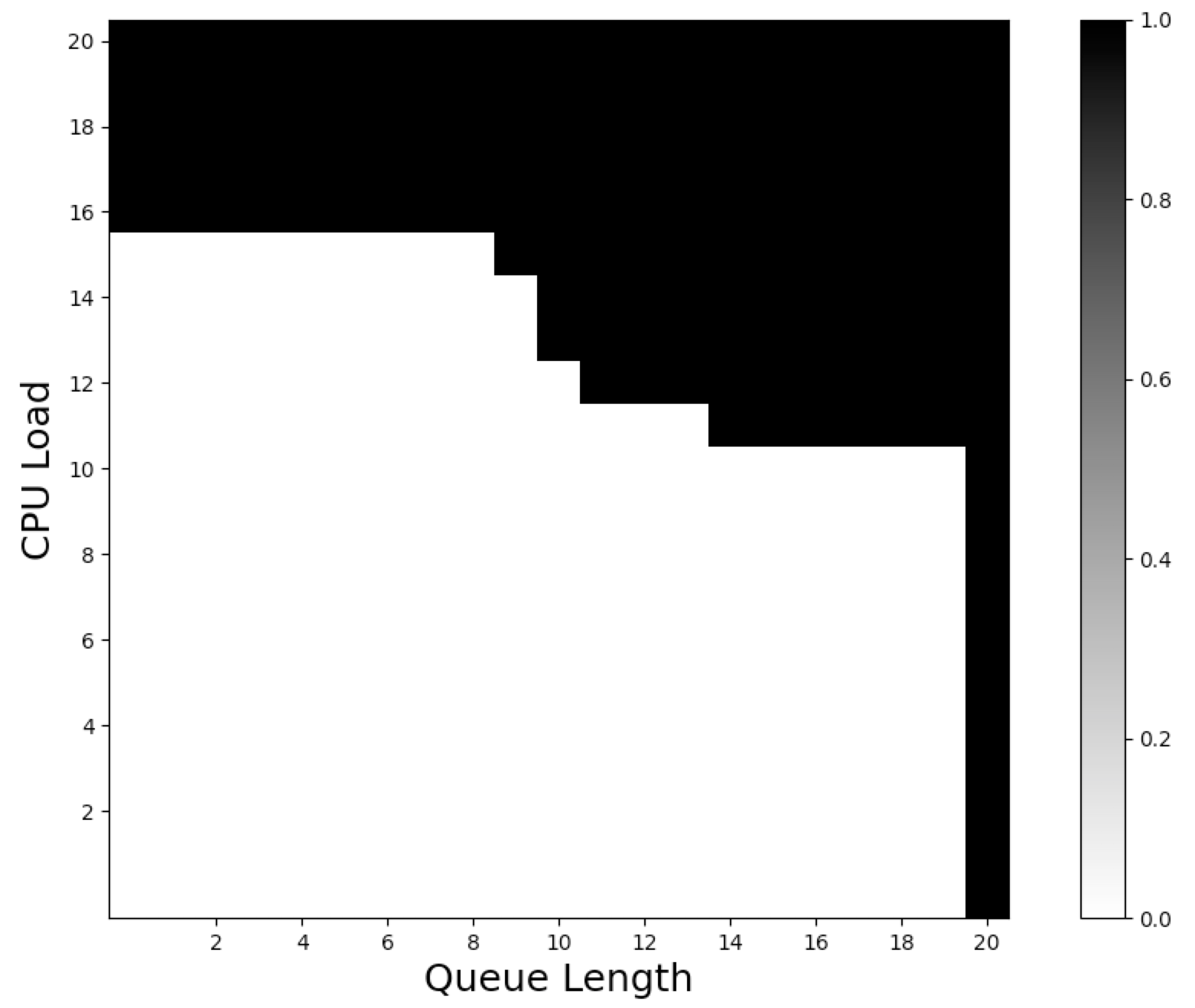}
      \caption{DP $\lambda = 6$}
      \label{fig:plan_l_12}
    \end{subfigure}%
    \hfill
    \begin{subfigure}[t]{0.5\linewidth}
      \centering
      \includegraphics[width=\textwidth]{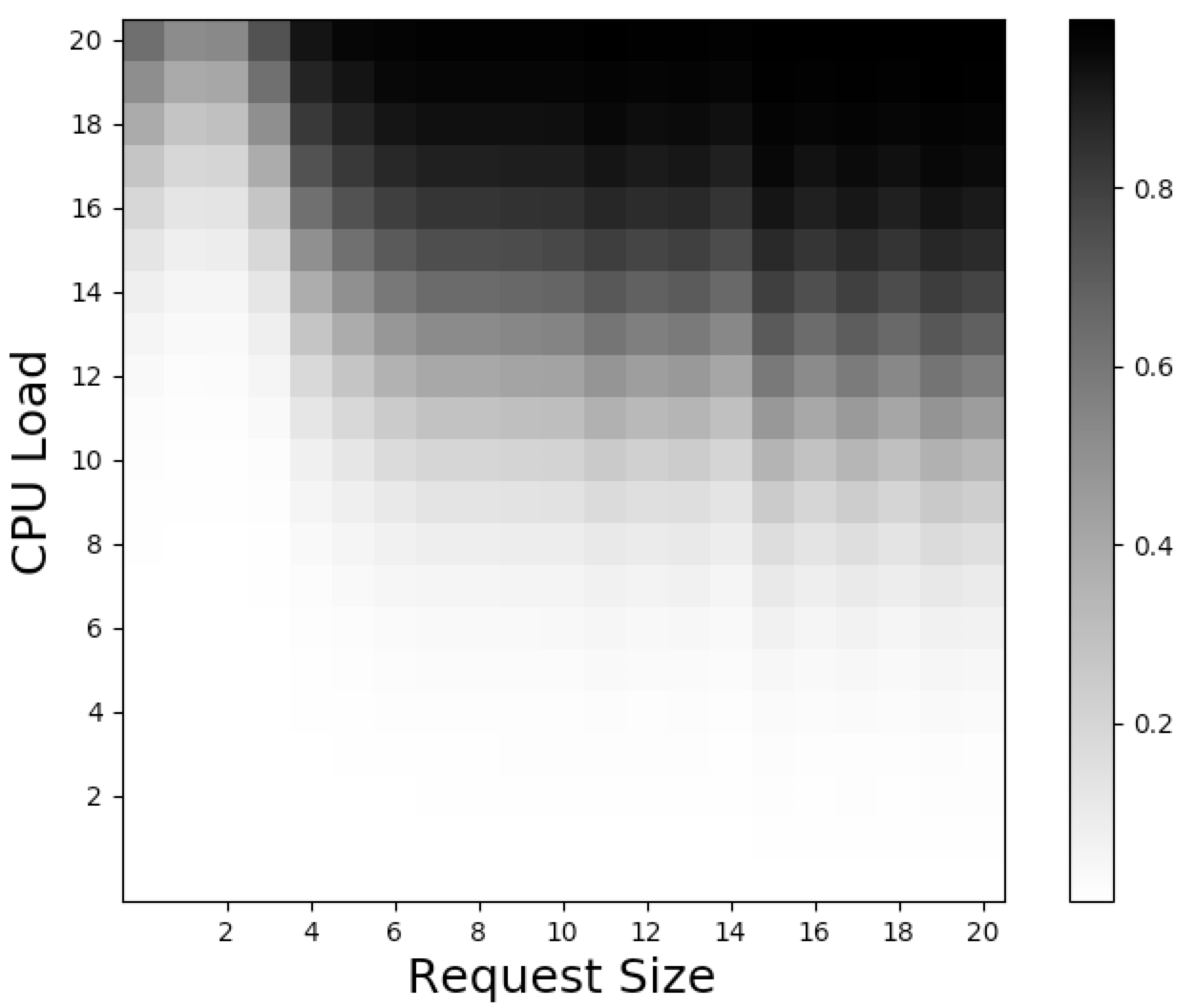}
      \caption{SALMUT $\lambda = 6$}
      \label{fig:salmut_l_12}
    \end{subfigure}%
    \caption{Comparing the optimal policy and converged policy of SALMUT along one of the sample paths. The colorbar represents the probability of the offloading action.}
    \label{fig:policy_structure}
\end{figure}

The plots for Scenario 1 (Fig.~\ref{fig:scenario1}) show that PPO converges to the optimal policy in less than $10^5$ steps, SALMUT and A2C takes around $2 \times 10^5$ steps, whereas Q-learning takes around $5 \times 10^5$ steps to converge. The policies for all the algorithms remain stable after convergence. 
Upon further analysis on the structure of the optimal policy, we observe that the structure of the optimal policy of SALMUT (Fig.~\ref{fig:salmut_l_12}) differs from that of the optimal policy computed using DP (Fig.~\ref{fig:plan_l_12}). There is a slight difference in the structure of these policies when buffer size (x) is low and CPU load ($\ell$) is high, which occurs because these states are reachable with a very low probability and hence SALMUT doesn't encounter these states in the simulation often to be able to learn the optimal policy in these states.
The plots from Scenario 2 (Fig.~\ref{fig:scenario2}) show similar behavior when $\lambda$ is constant. When $\lambda$ changes significantly, we observe all RL algorithms except Q-learning are able to adapt to the drastic but stable changes in the environment. Once the load stabilizes, all the algorithms are able to readjust to the changes and perform  close to the optimal policy. The plots from Scenario 3 (Fig.~\ref{fig:scenario3}) show similar behavior to Scenario 1, i.e. small but frequent changes in the environment do not impact the learning performance of reinforcement learning algorithms.

The plots from Scenario 4-6 (Fig.~\ref{fig:scenario4}-\ref{fig:scenario6}) show consistent performance with varying users. The RL algorithms including Q-learning show similar performance for most of the time-steps except in Scenario 5, which is similar to the behavior observed in Scenario 2. The Q-learning algorithm also performs poorly when the load suddenly changes in Scenarios 4 and 6. This could be due to the fact that Q-learning takes longer to adjust to a more aggressive offloading policy.

\subsection{Analysis of Training Time and Policy Interpretability}
The main difference among these four RL algorithms is the training time and
interpretability of policies. We ran our experiments on a server with Intel(R) Xeon(R) Gold
6148 CPU @ 2.40GHz processor. The training time of all the RL algoirthms is shown in
Table~\ref{table:train_time}. The mean training time is computed based on a single scenario over different runs and averaged across all the six scenarios. SALMUT is about 28 times faster to
train than PPO and 17 times faster than A2C. SALMUT does not require a non-linear function approximator such as Neural Networks (NN) to represent its policy, making the training time for SALMUT very fast. We observe that Q-learning is around 1.5 times faster than SALMUT as it does not need to update its policy parameters separately. Even though Q-learning is faster than SALMUT, Q-learning does not converge to an optimal policy when the request distribution changes. 

\begin{table}[!ht]
  \centering
  \caption{Training time of RL algorithms}
  \label{table:train_time}
  \begin{tabular}{@{}ccc@{}}
    \toprule
    \textbf{Algorithm} & \textbf{Mean Time (s)} & \textbf{Std-dev (s)}\\
    \midrule
    SALMUT & 98.23 & 4.86\\
    Q-learning & 62.73 & 1.57 \\
    PPO & 2673.17 & 23.33\\
    A2C & 1677.33 & 9.99\\
    \bottomrule
  \end{tabular}
\end{table}


By construction, SAL\-MUT searches for (randomized) threshold based policies.
For example, for Scenario~1, SALMUT converges to the policy shown in
Fig.~\ref{fig:salmut_l_12}.
It is easy for
a network operator to interpret such threshold based strategies and decide
whether to deploy them or not. In contrast, in deep RL algorithms such as PPO
and A2C, the policy is parameterized using a neural network and it is
difficult to visualize the learned weights of such a policy and decide
whether the resultant policy is reasonable. 
Thus, by leveraging on the threshold structure of the optimal policy, SALMUT
is able to learn faster and at the same time provide threshold based policies
which are easier to interpret.

The policy of SALMUT is completely characterized by the threshold vector $\tau$, making it storage efficient too. The threshold-nature of the optimal policy computed by SALMUT, can be easily interpreted by just looking at the threshold vector $\tau$ (see Fig.~\ref{fig:salmut_l_12}), making it easy to debug and estimate the behavior of the system operating under such policies. However, the policies learned by A2C and PPO are the learned weights of the NN, which are undecipherable and may lead to unpredictable results occasionally. It is very important that the performance of real-time systems be predictable and reliable, which has hindered the adoption of NNs in real-time deployments.  

\begin{figure*}[!t]
    \centering
    \begin{subfigure}[t]{0.32\linewidth}
      \centering
      \includegraphics[width=\textwidth]{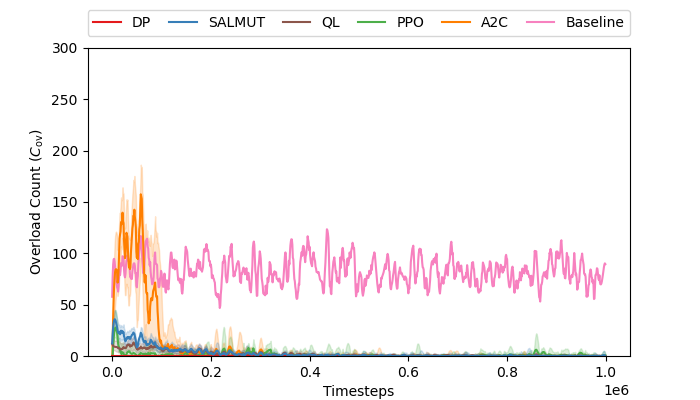}
      \caption{Scenario 1}
      \label{fig:ov_scenario1}
    \end{subfigure}%
    \hfill
    \begin{subfigure}[t]{0.32\linewidth}
      \centering
      \includegraphics[width=\textwidth]{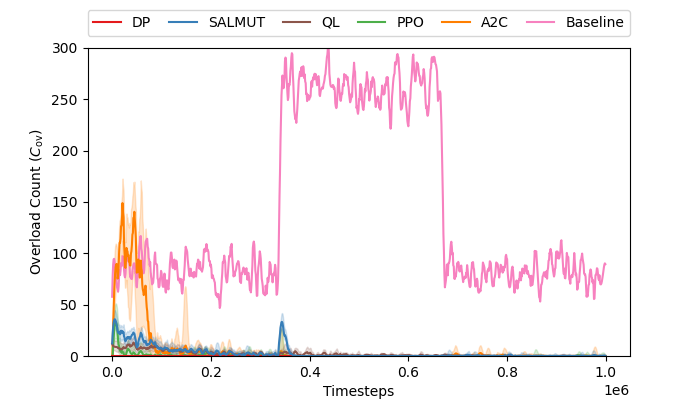}
      \caption{Scenario 2}
      \label{fig:ov_scenario2}
    \end{subfigure}%
    \hfill
    \begin{subfigure}[t]{0.32\linewidth}
      \centering
      \includegraphics[width=\textwidth]{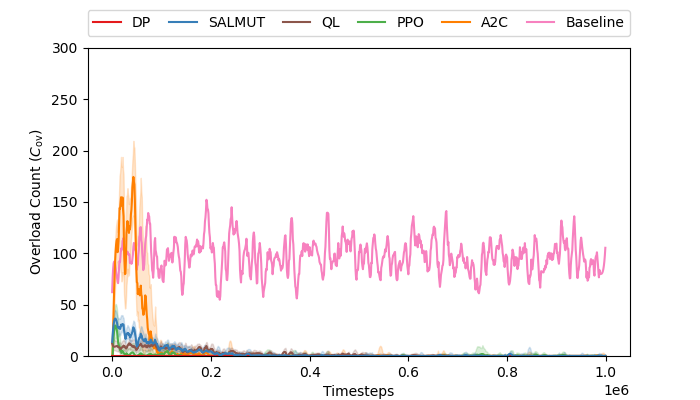}
      \caption{Scenario 3}
      \label{fig:ov_scenario3}
    \end{subfigure}%

    \begin{subfigure}[t]{0.32\linewidth}
      \centering
      \includegraphics[width=\textwidth]{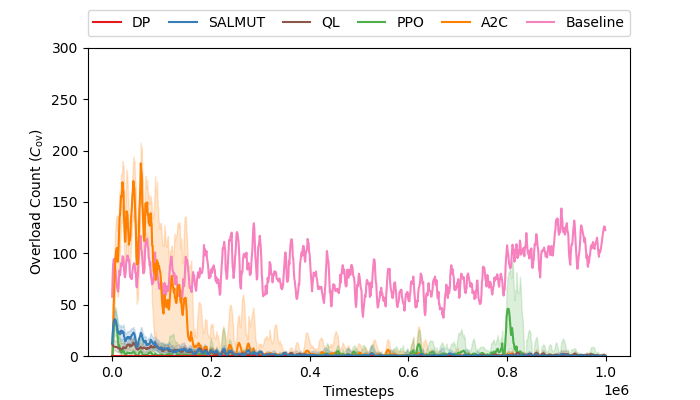}
      \caption{Scenario 4}
      \label{fig:ov_scenario4}
    \end{subfigure}%
    \hfill
    \begin{subfigure}[t]{0.32\linewidth}
      \centering
      \includegraphics[width=\textwidth]{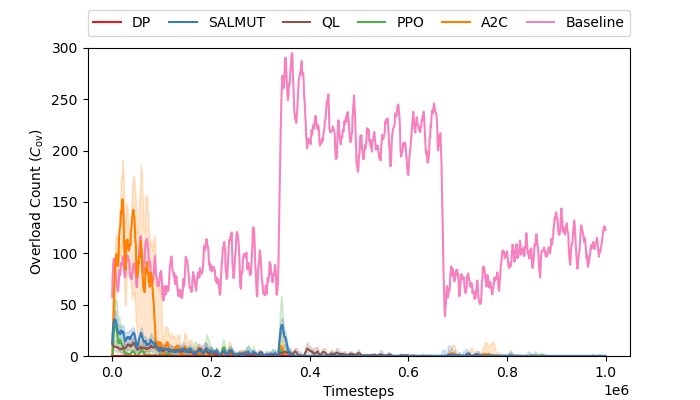}
      \caption{Scenario 5}
      \label{fig:ov_scenario5}
    \end{subfigure}%
    \hfill
    \begin{subfigure}[t]{0.32\linewidth}
      \centering
      \includegraphics[width=\textwidth]{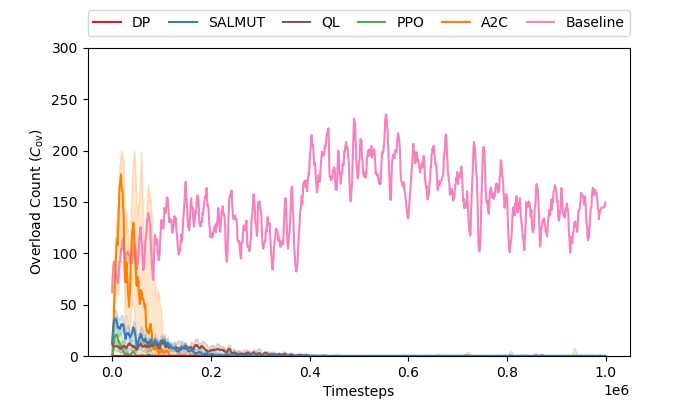}
      \caption{Scenario 6}
      \label{fig:ov_scenario6}
    \end{subfigure}%

    \caption{Comparing the number of times the system goes into the overloaded state at each evaluation step. The trajectory of event arrival and departure is fixed for all evaluation steps and across all algorithms for the same arrival distribution.} 
    \label{fig:overload}
\end{figure*}

\begin{figure*}[!t]
    \centering
    \begin{subfigure}[t]{0.32\linewidth}
      \centering
      \includegraphics[width=\textwidth]{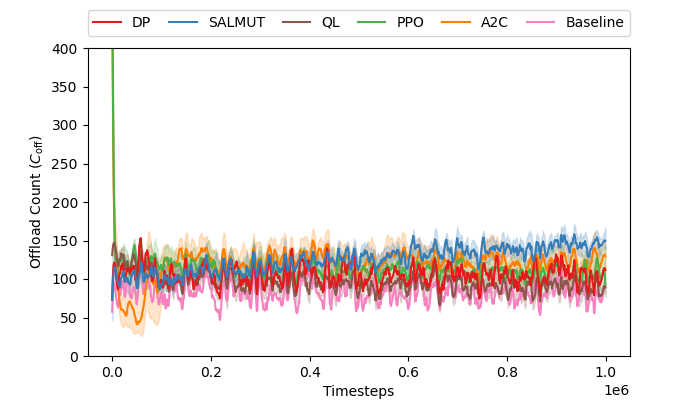}
      \caption{Scenario 1}
      \label{fig:of_scenario1}
    \end{subfigure}%
    \hfill
    \begin{subfigure}[t]{0.32\linewidth}
      \centering
      \includegraphics[width=\textwidth]{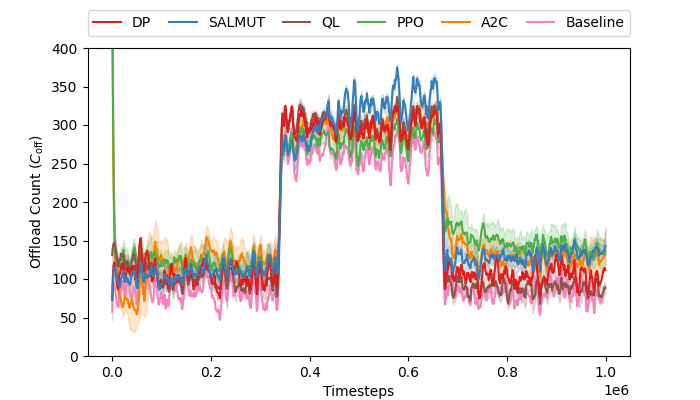}
      \caption{Scenario 2}
      \label{fig:of_scenario2}
    \end{subfigure}%
    \hfill
    \begin{subfigure}[t]{0.32\linewidth}
      \centering
      \includegraphics[width=\textwidth]{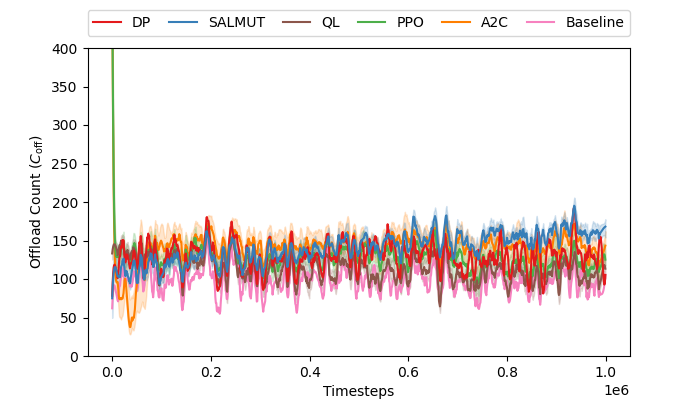}
      \caption{Scenario 3}
      \label{fig:of_scenario3}
    \end{subfigure}%

    \begin{subfigure}[t]{0.32\linewidth}
      \centering
      \includegraphics[width=\textwidth]{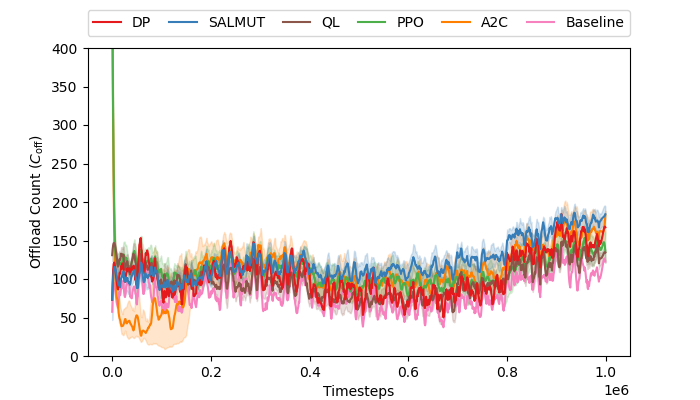}
      \caption{Scenario 4}
      \label{fig:of_scenario4}
    \end{subfigure}%
    \hfill
    \begin{subfigure}[t]{0.32\linewidth}
      \centering
      \includegraphics[width=\textwidth]{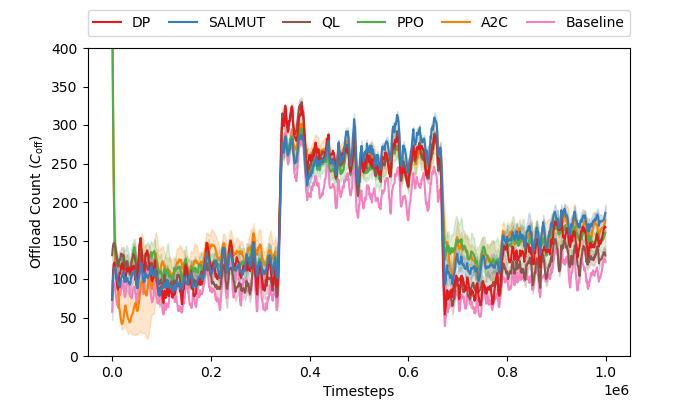}
      \caption{Scenario 5}
      \label{fig:of_scenario5}
    \end{subfigure}%
    \hfill
    \begin{subfigure}[t]{0.32\linewidth}
      \centering
      \includegraphics[width=\textwidth]{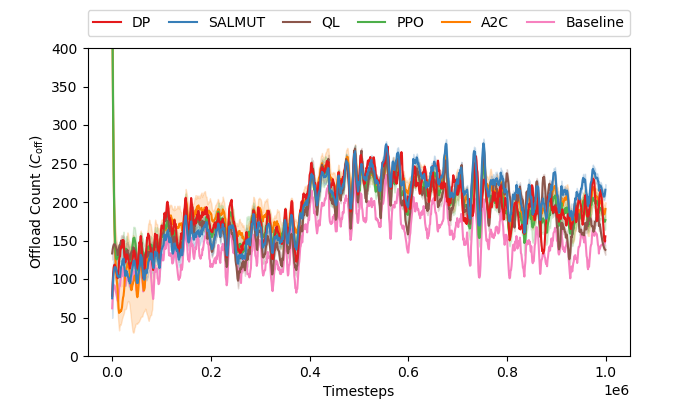}
      \caption{Scenario 6}
      \label{fig:of_scenario6}
    \end{subfigure}%

    \caption{Comparing the number of times the system performs offloading at each evaluation step. The trajectory of event arrival and departure is fixed for all evaluation steps and across all algorithms for the same arrival distribution.} 
    \label{fig:offload}
\end{figure*}

\subsection{Behavioral Analysis of Policies}

We performed further analysis on the behavior of the learned policy by observing the number of times the system enters into an overloaded state and offloads incoming request. Let us define $C_{\OV}$ to be the number of times the system enters into an overloaded state and $C_{\OFF}$ to be the number of times the system offloads requests for every 1000 steps of training iteration. We generated a set of $10^6$ random numbers between 0 and 1, defined by $z_t$, where $t$ is the step count. We use this set of random numbers to fix the trajectory of events (arrival or departure) for all the experiments in this section. Similar to the experiment in the previous section, the number of users $N$ and the arrival rate $\lambda$ are fixed for 1000 steps and evolve according to the scenarios described in Fig.~\ref{fig:evolution}. The event is set to arrival if $z_t$ is less than or equal to $\lambda_t / (\lambda_t + \min\{x_t,k\}\mu)$, and set to departure otherwise. These experiments were carried out during the training time for 10 different seeds. We plot the median of the number of times a system goes into an overloaded state (Fig.~\ref{fig:overload}) and the number of requests offloaded by the system (Fig.~\ref{fig:offload}) along with the uncertainty band from the first to the third quartile for every 1000 steps. 

We observe in Fig.~\ref{fig:overload}, that all the algorithms (SALMUT, Q-learning, PPO, A2C) learn not to enter into the overloaded state. As seen in the case of total discounted cost (Fig.~\ref{fig:results}), PPO learns it instantly, followed by SALMUT, Q-learning, and A2C. The observation is valid for all the different scenarios we tested. We observe that for Scenario-4, PPO enters the overloaded state at around $0.8 \times 10^6$ which is due to the fact the $\sum_i \lambda_i$ increases drastically at that point (seen in Fig.~\ref{fig:n_v_lambda_scenario4}) and we also see its effect on the cost in Fig.~\ref{fig:scenario4} at that time. We also observe that SALMUT enters into overloaded states when the request distribution changes drastically in Scenario 2 and 5. It is able to recover quickly and adapt its threshold policy to a more aggressive offloading policy. The baseline algorithms, on the other hand, enters into the overloaded state quiet often. 

We observe in Fig.~\ref{fig:offload}, that the algorithms (SALMUT, PPO, A2C) learn to adjust their offloading rate to avoid overloaded state. The number of times requests have been offloaded is directly proportional to the total arrival rate of all the users at that time. When the arrrival rate increases, the number of times the offloading occurs also increases in the interval. We see that even though the offloaded requests are higher for the RL algorithms than the baseline algorithm in all scenarios and timesteps, the difference between the number of times they offload is not significant implying that the RL algorithms learn policies that offload at the right moment as to not lead the system into an overloaded state. We perform further analysis of this behavior for the docker-testbed (see Fig.~\ref{fig:pareto_front}) and the results are similar for the simulations too.

\section{Testbed Implementation and Results}\label{sec:experiments-2}

\begin{figure}[!t]
 \centering
 \includegraphics[width=\linewidth]{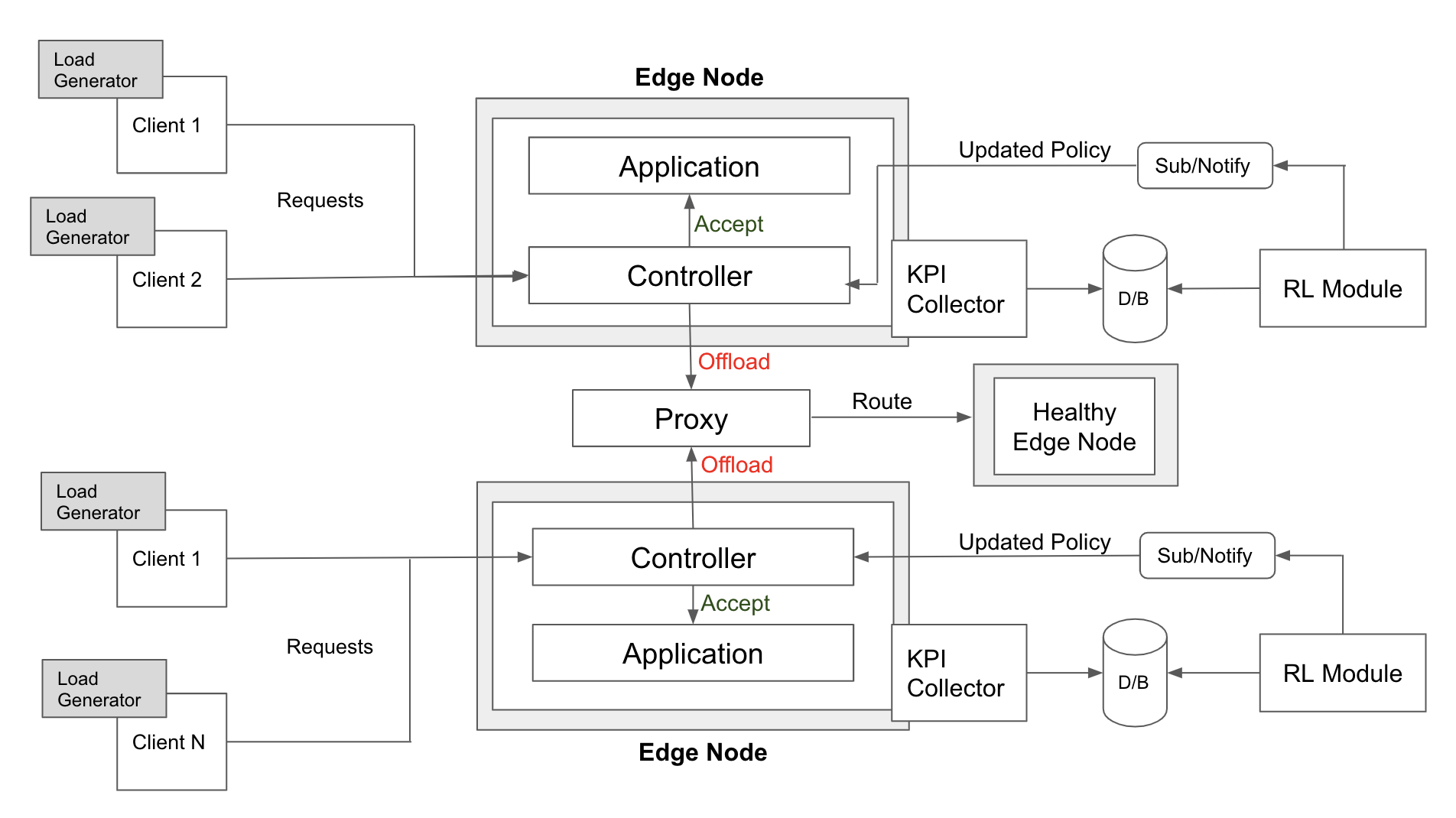}
 \caption{The overview of the docker-testbed environment.}
 \label{fig:testbed}
\end{figure}

We test our proposed algorithm on a testbed resembling the MEC architecture in Fig.~\ref{fig:model}, but without the core network and backend cloud server for simplicity. We consider an edge node which serves a single application. Both the edge nodes and clients are implemented as containerized environments in a virtual machine. The overview of the testbed is shown in Fig.~\ref{fig:testbed}. The load generator generates requests for each client independently according to a time-varying Poisson process. The requests at the edge node are handled by the controller which decides either to accept the request or offload the request based on the policy for the current state of the edge node. If the action is "accept", the request is added to the request queue of the edge node, otherwise the request is offloaded to another healthy edge node via the proxy network. The Key Performance Indicator (KPI) collector copies the KPI metrics into a database at regular intervals. The RL modules uses these metrics to update its policies. The Subscriber/Notification (Sub/Notify) module notifies the controller about the updated policy. The controller now uses the updated policy to serve all future requests.

In our implementation, the number of clients $N$ served by an edge node and the request rate of the clients $\lambda$ is constant for at-least $100$ seconds. We define a step to be the execution of the testbed for $100$ seconds. Each request runs a workload on the edge node and consumes CPU resources $R$, where $R$ is a random variable. The states, actions, costs, next states for each step are stored in a buffer in the edge node. After the completion of a step, the KPI collector copies these buffers into a database. The RL module is then invoked, which loads its most recent policy and other parameters, and trains on this new data to update its policy. Once the updated policy is generated, it is copied in the edge node and is used by the controller for serving the requests for the next step.

We run our experiments for a total of $1000$ steps, where $N$ and $\lambda$ evolve according to Fig.~\ref{fig:evolution} for different scenarios, similar to the previous set of experiments. We consider an edge server with buffer size
$\mathsf{X} = 20$, CPU capacity $\mathsf{L} = 20$, $k = 2$ cores, service-rate $\mu = 3.0$
for each core, holding cost $h = 0.12$. The CPU capacity is discretized into
$20$ states for utilization $0-100 \%$, similar to the previous experiment. The CPU running cost is $c(\ell) = 30$ for $\ell \ge 18$, $c(\ell) = -0.2$ for $6 \le \ell \le 17$, and $c(\ell) = 0$ otherwise. The offload penalty is $p = 1$ for $\ell \ge 3$ and $p = 10$ for $\ell < 3$. We assume that the discrete time discount factor $\beta = \alpha/(\alpha + \nu)$ equals $0.99$.

\subsection{Results}

\begin{figure*}[!t]
    \centering
    \begin{subfigure}[t]{0.33\linewidth}
      \centering
      \includegraphics[width=\textwidth]{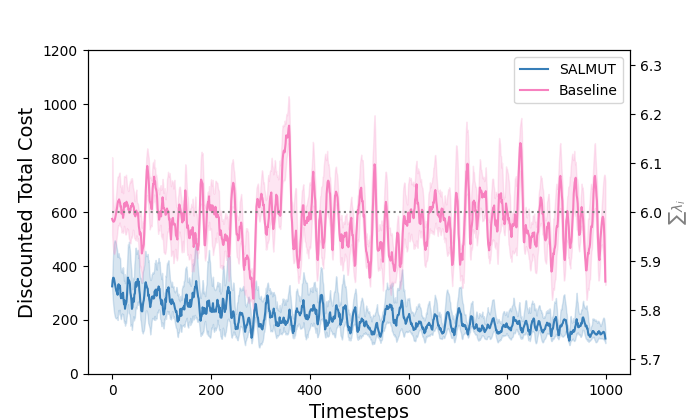}
      \caption{Scenario 1}
      \label{fig:scenario1_t}
    \end{subfigure}%
    \hfill
    \begin{subfigure}[t]{0.33\linewidth}
      \centering
      \includegraphics[width=\textwidth]{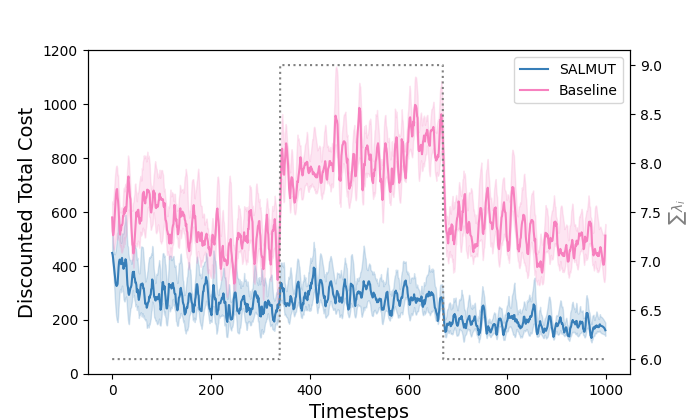}
      \caption{Scenario 2}
      \label{fig:scenario2_t}
    \end{subfigure}%
    \hfill
    \begin{subfigure}[t]{0.33\linewidth}
      \centering
      \includegraphics[width=\textwidth]{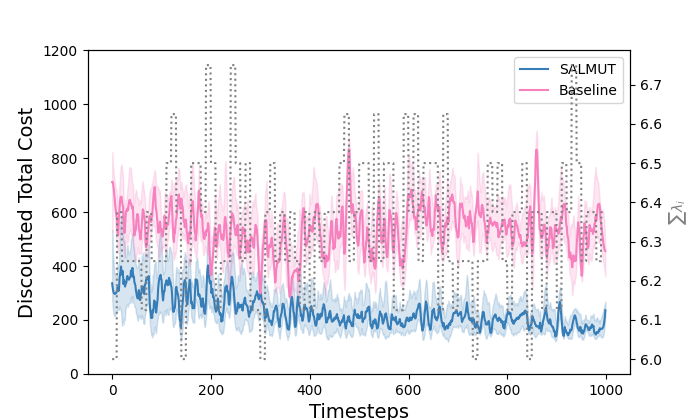}
      \caption{Scenario 3}
      \label{fig:scenario3_t}
    \end{subfigure}%

    \begin{subfigure}[t]{0.33\linewidth}
      \centering
      \includegraphics[width=\textwidth]{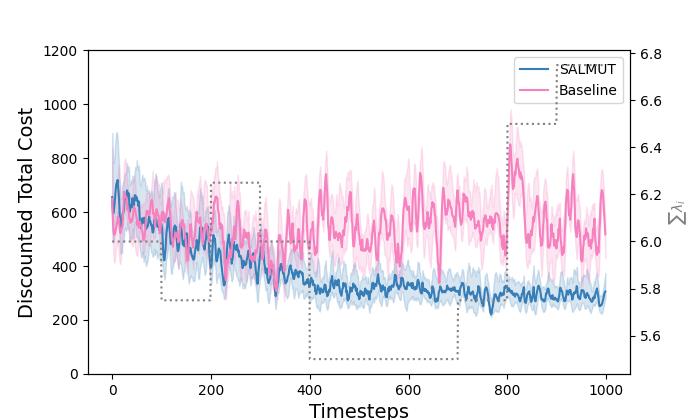}
      \caption{Scenario 4}
      \label{fig:scenario4_t}
    \end{subfigure}%
    \hfill
    \begin{subfigure}[t]{0.33\linewidth}
      \centering
      \includegraphics[width=\textwidth]{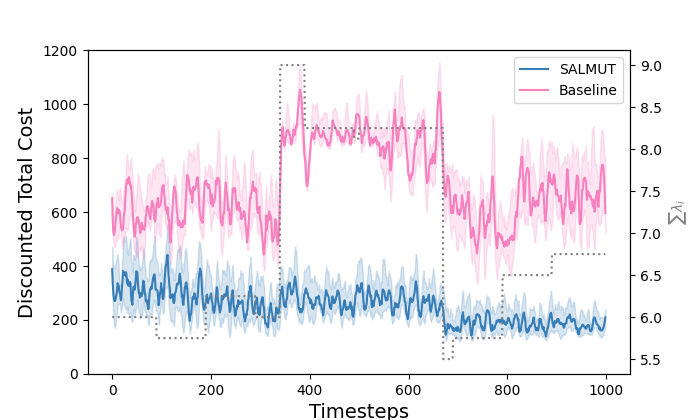}
      \caption{Scenario 5}
      \label{fig:scenario5_t}
    \end{subfigure}%
    \hfill
    \begin{subfigure}[t]{0.33\linewidth}
      \centering
      \includegraphics[width=\textwidth]{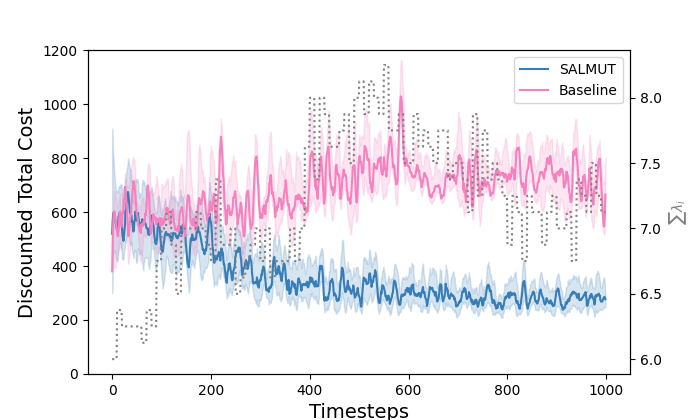}
      \caption{Scenario 6}
      \label{fig:scenario6_t}
    \end{subfigure}%
    \caption{Performance of RL algorithms for different scenarios in the end-to-end testbed we created. We also plot the total request arrival rate ($\sum_i \lambda_i$) on the right-hand side of y-axis in gray dotted lines.} 
    \label{fig:results_testbed}
\end{figure*}

We run the experiments for SALMUT and baseline algorithm for a total of 1000 steps. We do not run the simulations for PPO and A2C algorithms in our testbed as these algorithms cannot be trained in real-time because the time they require to process each sample is more than the sampling interval. The performance of SALMUT and baseline algorithm is evaluated at every step by computing the discounted total cost for that step using the cost buffers which are stored in the database. The experiment is repeated $5$ times and the median performance with an uncertainty band from the first to the third quartile are plotted in Fig.~\ref{fig:results_testbed} along with the total request arrival rate ($\sum_i \lambda_i$) in gray dotted lines.

For Scenario~1 (Fig.~\ref{fig:scenario1_t}), we observe that the SALMUT algorithm outperforms the baseline algorithm right from the start, indicating that SALMUT updates its policy swiftly at the start and slowly converges towards optimal performance, whereas the baseline incurs high cost throughout. Since SALMUT policies converge towards optimal performance after some time, they
are also able to adapt quickly in Scenarios~2--6 (Fig.~\ref{fig:scenario2_t}-\ref{fig:scenario6_t}) and keep track of the
time-varying arrival rates and number of users. We observe that SALMUT takes some time to learn a good policy, but once it learns the policy, it adjusts to frequent but small changes in $\lambda$ and $N$ very well (see Fig.~\ref{fig:scenario3_t} and \ref{fig:scenario4_t}). If the request rate changes drastically, the performance decreases a little (which is bound to happen as the total requests to process are much larger than the server's capacity) but the magnitude of the performance drop is much lesser in SALMUT as compared to the baseline, seen in Fig.~\ref{fig:scenario2_t}, \ref{fig:scenario5_t} and \ref{fig:scenario6_t}. It is because the baseline algorithms incur high overloading cost for these requests whereas SALMUT incurs offloading costs for the same requests. Further analysis on this is present in Section 4.2.2.

\begin{figure*}[!t]
    \centering
    \begin{subfigure}[t]{0.33\linewidth}
      \centering
      \includegraphics[width=\textwidth]{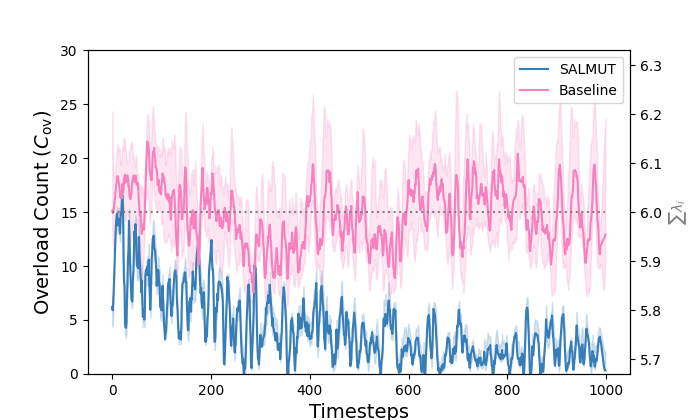}
      \caption{Scenario 1}
      \label{fig:ov_scenario1_t}
    \end{subfigure}%
    \hfill
    \begin{subfigure}[t]{0.33\linewidth}
      \centering
      \includegraphics[width=\textwidth]{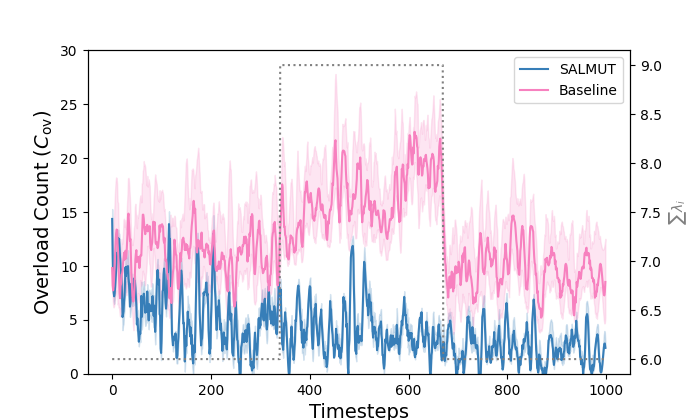}
      \caption{Scenario 2}
      \label{fig:ov_scenario2_t}
    \end{subfigure}%
    \hfill
    \begin{subfigure}[t]{0.33\linewidth}
      \centering
      \includegraphics[width=\textwidth]{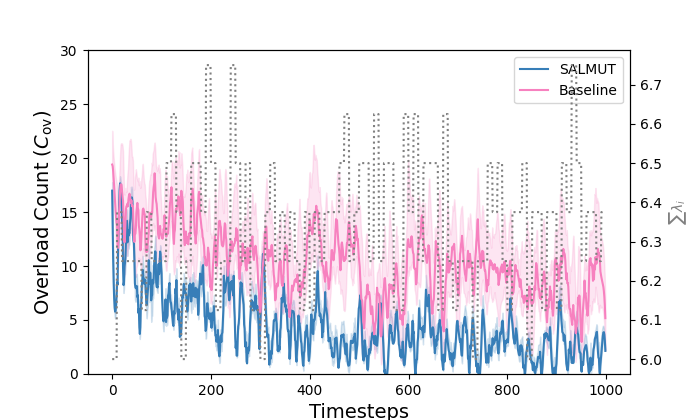}
      \caption{Scenario 3}
      \label{fig:ov_scenario3_t}
    \end{subfigure}%

    \begin{subfigure}[t]{0.33\linewidth}
      \centering
      \includegraphics[width=\textwidth]{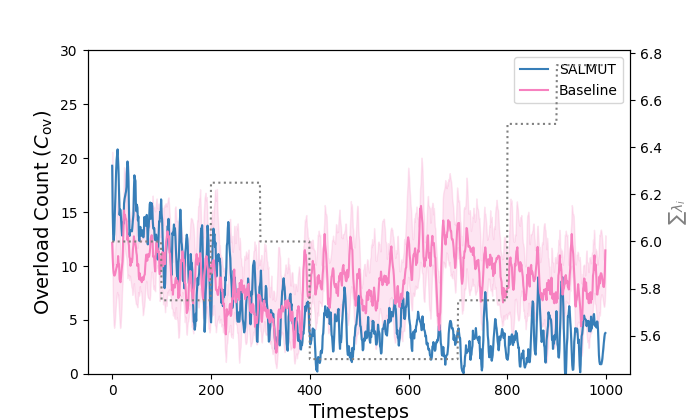}
      \caption{Scenario 4}
      \label{fig:ov_scenario4_t}
    \end{subfigure}%
    \hfill
    \begin{subfigure}[t]{0.33\linewidth}
      \centering
      \includegraphics[width=\textwidth]{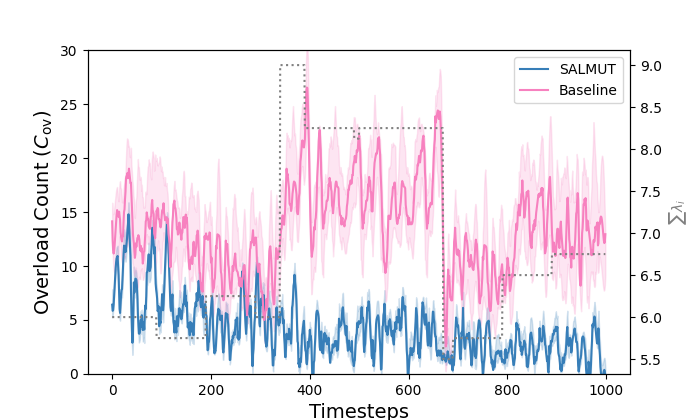}
      \caption{Scenario 5}
      \label{fig:ov_scenario5_t}
    \end{subfigure}%
    \hfill
    \begin{subfigure}[t]{0.33\linewidth}
      \centering
      \includegraphics[width=\textwidth]{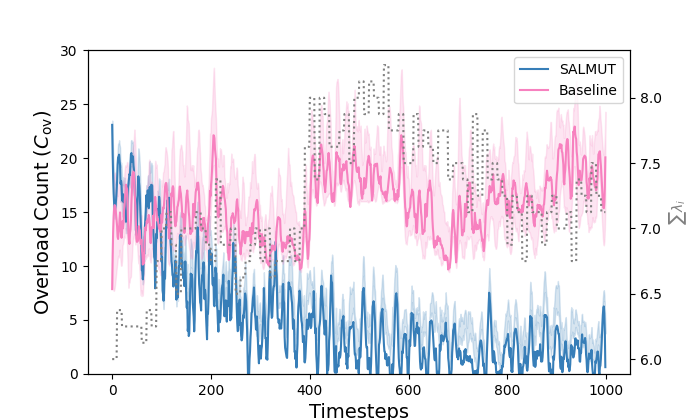}
      \caption{Scenario 6}
      \label{fig:ov_scenario6_t}
    \end{subfigure}%

    \caption{Comparing the number of times the system goes into the overloaded state at each step in the end-to-end testbed we created. We also plot the total request arrival rate ($\sum_i \lambda_i$) on the right-hand side of y-axis in gray dotted lines.} 
    \label{fig:overload_testbed}
\end{figure*}

\begin{figure*}[!t]
    \centering
    \begin{subfigure}[t]{0.33\linewidth}
      \centering
      \includegraphics[width=\textwidth]{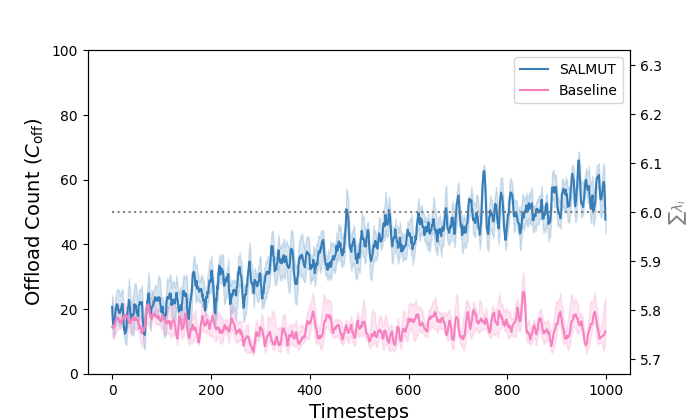}
      \caption{Scenario 1}
      \label{fig:of_scenario1_t}
    \end{subfigure}%
    \hfill
    \begin{subfigure}[t]{0.33\linewidth}
      \centering
      \includegraphics[width=\textwidth]{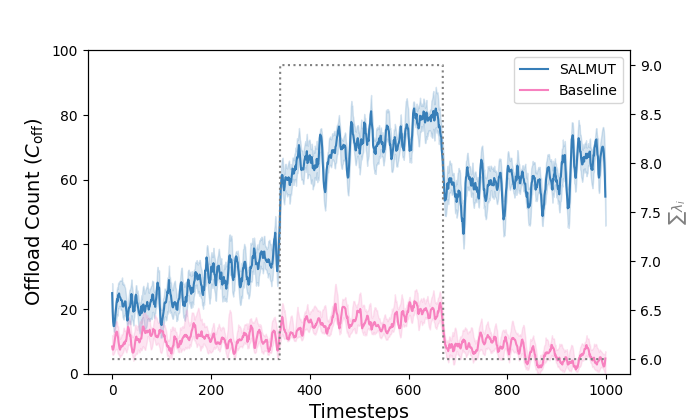}
      \caption{Scenario 2}
      \label{fig:of_scenario2_t}
    \end{subfigure}%
    \hfill
    \begin{subfigure}[t]{0.33\linewidth}
      \centering
      \includegraphics[width=\textwidth]{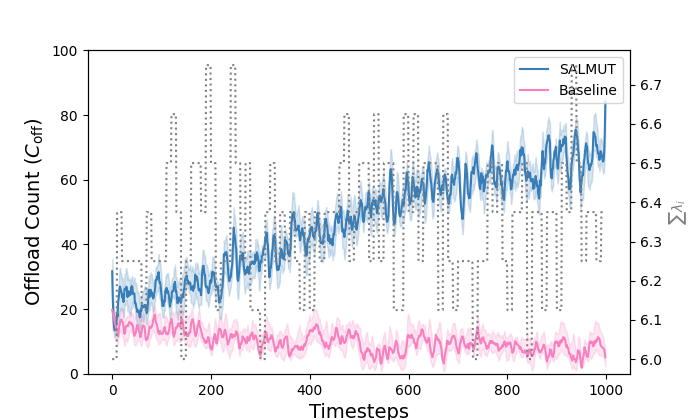}
      \caption{Scenario 3}
      \label{fig:of_scenario3_t}
    \end{subfigure}%
   
    \begin{subfigure}[t]{0.33\linewidth}
      \centering
      \includegraphics[width=\textwidth]{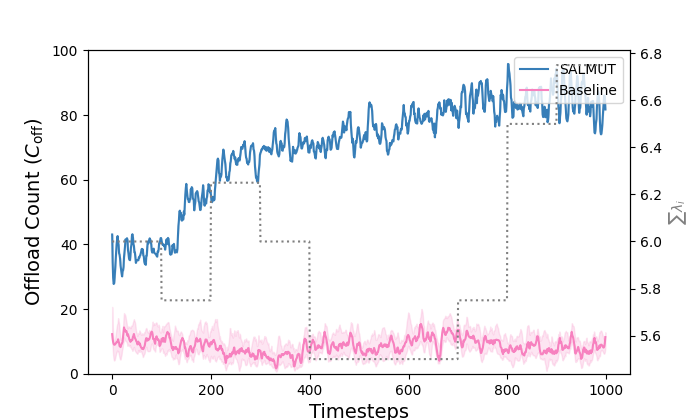}
      \caption{Scenario 4}
      \label{fig:of_scenario4_t}
    \end{subfigure}%
    \hfill
    \begin{subfigure}[t]{0.33\linewidth}
      \centering
      \includegraphics[width=\textwidth]{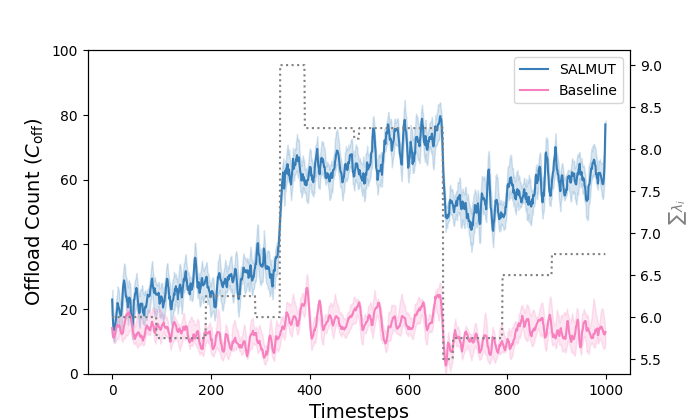}
      \caption{Scenario 5}
      \label{fig:of_scenario5_t}
    \end{subfigure}%
    \hfill
    \begin{subfigure}[t]{0.33\linewidth}
      \centering
      \includegraphics[width=\textwidth]{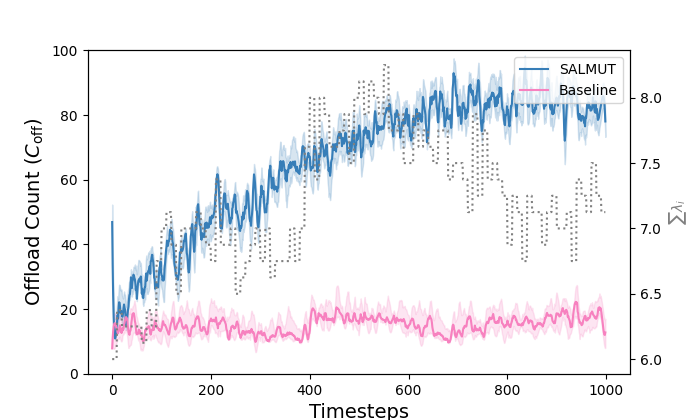}
      \caption{Scenario 6}
      \label{fig:of_scenario6_t}
    \end{subfigure}%

    \caption{Comparing the number of times the system performs offloading at each step in the end-to-end testbed we created. We also plot the total request arrival rate ($\sum_i \lambda_i$) on the right-hand side of y-axis in gray dotted lines.} 
    \label{fig:offload_testbed}
\end{figure*}

\begin{figure*}[!t]
    \centering
    \begin{subfigure}[t]{0.33\linewidth}
      \centering
      \includegraphics[width=\textwidth]{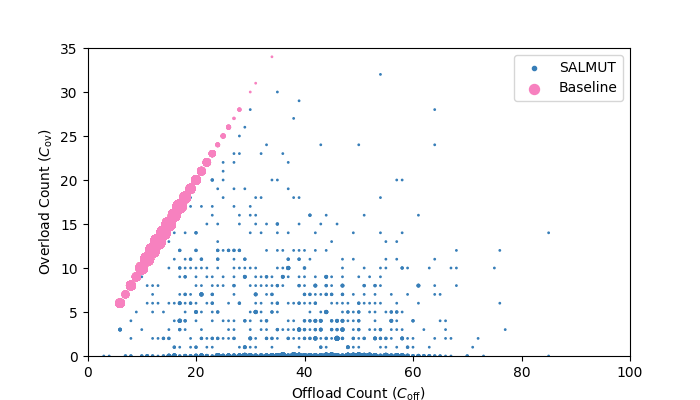}
      \caption{Scenario 1}
      \label{fig:par_scenario1_t}
    \end{subfigure}%
    \hfill
    \begin{subfigure}[t]{0.33\linewidth}
      \centering
      \includegraphics[width=\textwidth]{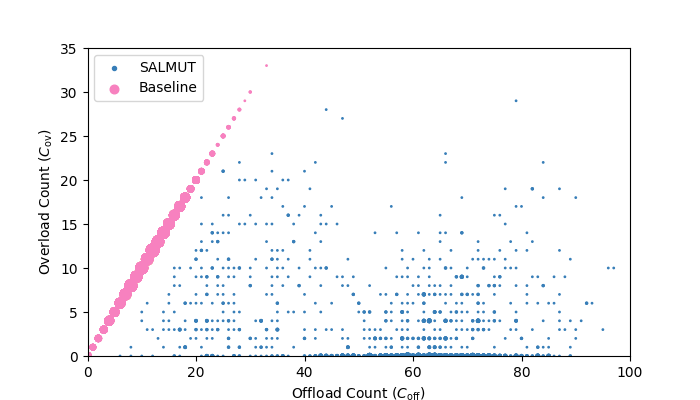}
      \caption{Scenario 2}
      \label{fig:par_scenario2_t}
    \end{subfigure}%
    \hfill
    \begin{subfigure}[t]{0.33\linewidth}
      \centering
      \includegraphics[width=\textwidth]{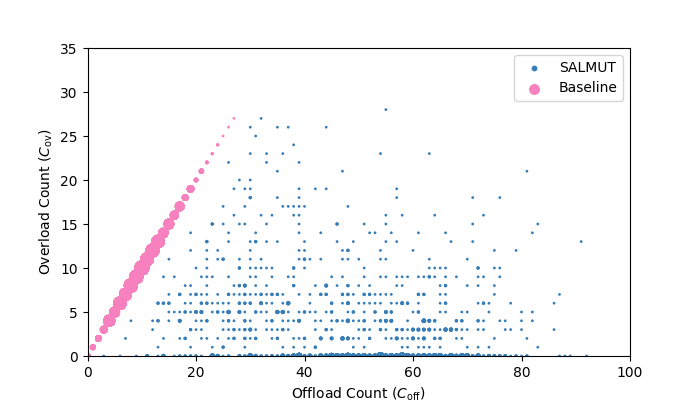}
      \caption{Scenario 3}
      \label{fig:par_scenario3_t}
    \end{subfigure}%
   
    \begin{subfigure}[t]{0.33\linewidth}
      \centering
      \includegraphics[width=\textwidth]{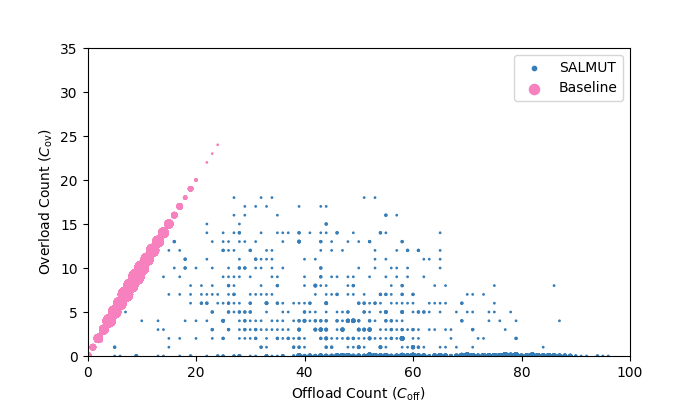}
      \caption{Scenario 4}
      \label{fig:par_scenario4_t}
    \end{subfigure}%
    \hfill
    \begin{subfigure}[t]{0.33\linewidth}
      \centering
      \includegraphics[width=\textwidth]{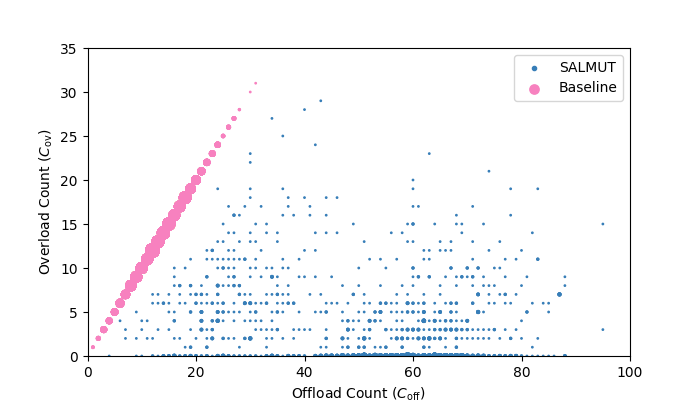}
      \caption{Scenario 5}
      \label{fig:par_scenario5_t}
    \end{subfigure}%
    \hfill
    \begin{subfigure}[t]{0.33\linewidth}
      \centering
      \includegraphics[width=\textwidth]{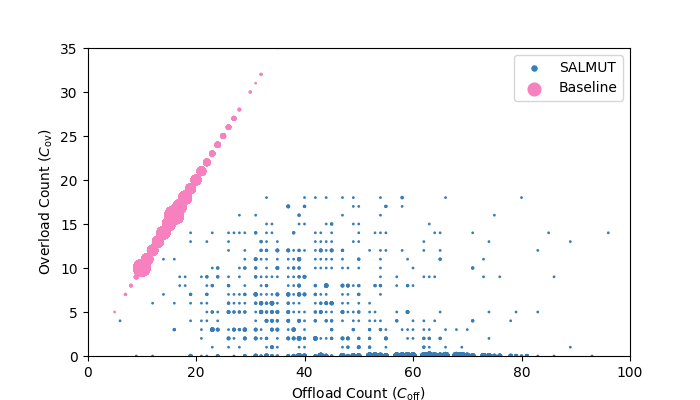}
      \caption{Scenario 6}
      \label{fig:par_scenario6_t}
    \end{subfigure}%

    \caption{Scatter-plot of $C_{\OV}$ Vs $C_{\OFF}$ for SALMUT and baseline algorithm in the end-to-end testbed we created. The width of the points is proportional to its frequency.} 
    \label{fig:pareto_front}
\end{figure*}

\subsection{Behavioral Analysis}

We perform behavior analysis of the learned policy by observing the number of times the system enters into an overloaded state (denoted by $C_{\OV}$) and the number of incoming request offloaded by the edge node (denoted by $C_{\OFF}$) in a window of size 100. These plots are shown in Fig.~\ref{fig:overload_testbed} \& \ref{fig:offload_testbed}.

We observe from Fig.~\ref{fig:overload_testbed} that the number of times the edge node goes into an overload state while following policy executed by SALMUT is much less than the baseline algorithm. Even when the system goes into an overloaded state, it is able to recover quickly and does not suffer from performance deterioration. From Fig.~\ref{fig:of_scenario2_t} and \ref{fig:of_scenario5_t} we can observe that in Scenarios 2 and 5, when the request load increases drastically (at around 340 steps), $C_{\OFF}$ increases and its effects can also be seen in the overall discounted cost in Fig.~\ref{fig:scenario2_t} and \ref{fig:scenario5_t} at around the same time. SALMUT is able to adapt its policy quickly and recover quickly. We observe in Fig.~\ref{fig:offload_testbed} that SALMUT performs more offloading as compared to the baseline algorithm.

A policy that offloads often and does not go into an overloaded state may not necessarily minimize the total cost. We did some further investigation by visualizing the scatter-plot (Fig.~\ref{fig:pareto_front}) of the overload count ($C_{\OV}$) on the y-axis and the offload count ($C_{\OFF}$) on the x-axis for both SALMUT and the baseline algorithm for all the scenarios described in Fig.~\ref{fig:evolution}. We observe that SALMUT keeps $C_{\OV}$ much lower than the baseline algorithm at the cost of increased $C_{\OFF}$. We observe from Fig.~\ref{fig:pareto_front} that the slope for the plot is linear for baseline algorithms because they are offloading reactively. SALMUT, on the other hand, learns a behavior that is analogous to pro-active offloading, where it benefits from the offloading action it takes by minimizing $C_{\OV}$.


\section{Conclusion and Limitations} \label{sec:conclusion}
In this paper we considered a single node optimal policy for overload protection on the edge server in
a time varying environment. We proposed a RL-based adaptive low-complexity admission control policy that exploits the structure of the optimal policy and finds a policy that is easy to interpret. Our proposed algorithm performs as well as the standard deep RL algorithms but has a better computational and storage complexity, thereby significantly reducing the total training time. Therefore, our proposed algorithm is more suitable for deployment in real systems for online training. 

The results presented in this paper can be extended in several directions. In addition to CPU overload, one could consider other resource bottlenecks such as disk I/O, RAM utilization, etc. It may be desirable to simultaneously consider multiple resource constraints. Along similar lines, one could consider multiple applications with different resource requirements and different priority. If it can be established that the optimal policy in these more sophisticated setup has a threshold structure similar to Proposition \ref{prop:policy}, then we can apply the framework developed in this paper.


The discussion in this paper was restricted to a single node. These results could also provide a foundation to investigate node overload protection in multi-node clusters where additional challenges such as routing, link failures, and network topology shall be considered. 

\IEEEpeerreviewmaketitle

\appendices
\section{Proof of Proposition 1}
\label{appendix:A}
Let $\delta(x) = \lambda / (\lambda + \min(k,x)\mu)$. We define a sequence of value functions $\{V_n\}_{n \ge 0}$ as follows
\setlength{\belowdisplayskip}{0pt}
\setlength{\belowdisplayshortskip}{0pt}
\[
    V_0(x,\ell) = 0
\] and for $n \ge 0$
\[
    V_{n+1}(x, \ell) = \min\{Q_{n+1}(x,\ell, \ACCEPT), Q_{n+1}(x, \ell, \REJECT)\},
\]
\setlength{\belowdisplayskip}{5pt}
\setlength{\belowdisplayshortskip}{5pt}
where 
  \begin{align*}
    Q_{n+1}(x,&\ell,\ACCEPT) = \frac{1}{\alpha + \nu} \bigl[ h[x-k]^{+} + c(\ell) \bigr]
    \notag \\
    &  
    + \beta \bigg[
      \delta(x) \sum_{r = 1}^{\mathsf{R}}P(r)
      V_n([x+1]_{\mathsf{X}}, [\ell + r]_{\mathsf{L}})
    \notag \\
    & \qquad + 
      (1 - \delta(x)) \sum_{r = 1}^{\mathsf{R}}P(r)
    V_n([x-1]^{+}, [\ell - r]^{+}) \biggr]
    \shortintertext{and}
    Q_{n+1}(x,&\ell,\REJECT) = \frac{1}{\alpha + \nu} \bigl[ h[x-k]^{+} + c(\ell) + p(\ell)  \bigr]
    \notag \\
    &  
    + \beta 
      (1 - \delta(x)) \sum_{r = 1}^{\mathsf{R}}P(r)
      V_n([x-1]^{+}, [\ell - r]^{+}),
  \end{align*}
  where $[x]_{\mathsf{B}}$ denotes $\min\{x,\mathsf{B}\}$. 

Note that $\{V_n\}_{n \ge 0}$ denotes the iterates of the value iteration algorithm, and from \cite{Puterman1994}, we know that
\begin{equation}
    \label{eq:fixed_point}
    \lim_{n \to \infty} V_n(x, \ell) = V(x,\ell), \quad \forall x, \ell
\end{equation}
where $V$ is the unique fixed point of (\ref{eq:DP}).

We will show that (see Lemma \ref{lemma:value_prop} below) each $V_n(x, \ell)$ satisfies the property of Proposition \ref{prop:value}. Therefore, by (\ref{eq:fixed_point}) we get that $V$ also satisfies the property.

\begin{lemma}
    \label{lemma:value_prop}
    For each $n \ge 0$ and $x \in \{0, ..., X\}$, $V_n(x, \ell)$ is weakly increasing in $\ell$.
\end{lemma}
\begin{proof}
We prove the result by induction. Note that $V_0(x,\ell) = 0$ and is trivially weakly increasing in $\ell$. This forms the basis of the induction. Now assume that $V_n(x,\ell)$ is weakly increasing in $\ell$. Consider iteration $n+1$. Let $x \in \{0, ..., X\}$ and $\ell_1, \ell_2 \in \{0, ..., L\}$ such that $\ell_1 < \ell_2$. Then,
\newcommand\numberthis{\addtocounter{equation}{1}\tag{\theequation}}
\begin{align*}
    Q_{n+1}(x,&\ell_1,\ACCEPT) = \frac{1}{\alpha + \nu} \bigl[ h[x-k]^{+} + c(\ell_1) \bigr]
    \notag \\
    & \qquad 
    + \beta \bigg[
      \delta(x) \sum_{r = 1}^{\mathsf{R}}P(r)
      V_n([x+1]_{\mathsf{X}}, [\ell_1 + r]_{\mathsf{L}})
    \notag \\
    & \qquad +
      (1 - \delta(x)) \sum_{r = 1}^{\mathsf{R}}P(r)
    V_n([x-1]^{+}, [\ell_1 - r]^{+}) \biggr] \\
    & \stackrel{(a)}{\le} \frac{1}{\alpha + \nu} \bigl[ h[x-k]^{+} + c(\ell_2) \bigr]
    \notag \\
    &  \qquad
    + \beta \bigg[
      \delta(x) \sum_{r = 1}^{\mathsf{R}}P(r)
      V_n([x+1]_{\mathsf{X}}, [\ell_2 + r]_{\mathsf{L}})
    \notag \\
    & \qquad + 
      (1 - \delta(x)) \sum_{r = 1}^{\mathsf{R}}P(r)
    V_n([x-1]^{+}, [\ell_2 - r]^{+}) \biggr] 
    \notag \\
    & = Q_{n+1}(x, \ell_2, \ACCEPT), \numberthis \label{eq:appA_1}
  \end{align*}
where $(a)$ follows from the fact that $c(\ell)$ and $V_n(x,\ell)$ are weakly increasing in $\ell$.
  
 By a similar argument, we can show that 
 \begin{equation}
    \label{eq:appA_2}
     Q_{n+1}(x,\ell_1, \REJECT) \le Q_{n+1}(x, \ell_2, \REJECT).
 \end{equation}
 
 Now,
 \begin{align*}
     & V_{n+1}(x,\ell_1) = \min\{Q_{n+1}(x,\ell_1, \ACCEPT), Q_{n+1}(x, \ell_1, \REJECT)\} \\
     & \qquad \stackrel{(b)}{\le} \min\{Q_{n+1}(x,\ell_2, \ACCEPT), Q_{n+1}(x, \ell_2, \REJECT)\} \\
     & \qquad = V_{n+1}(x, \ell_2),  \numberthis \label{eq:appA_3}
\end{align*}
where $(b)$ follows from (\ref{eq:appA_1}) and (\ref{eq:appA_2}). Eq. (\ref{eq:appA_3}) shows that $V_{n+1}(x,\ell)$ is weakly increasing in $\ell$. This proves the induction step. Hence, the result holds for the induction.

\end{proof}

\section{Proof of Proposition 2}
\label{appendix:B}
\begin{proof}
  Let $\delta(x) = \lambda/(\lambda + \min\{x,k\} \mu)$. Consider 
  \begin{align*}
    \Delta Q{(x,\ell)} = &\coloneqq
    Q(x,\ell,\REJECT) - Q(x,\ell,\ACCEPT) \\
    &=
    - \beta 
    \delta(x) \sum_{r = 1}^{\mathsf{R}}P(r)
      V([x]_{\mathsf{X}}, [\ell + r]_{\mathsf{L}})
    - p.
  \end{align*}
  For a fixed $x$, by Proposition~\ref{prop:value}, $\Delta Q(x,\ell)$ is
  weakly decreasing in~$\ell$. If it is optimal to reject a request at state
  $(x,\ell)$ (i.e., $\Delta Q(x,\ell) \le 0$), then for any $\ell' > \ell$, 
  \[
    \Delta Q(x,\ell') \le \Delta Q(x,\ell) \le 0;
  \]
  therefore, it is optimal to reject the request.

\end{proof}

\section{Proof of Optimality of SALMUT}
\label{appendix:C}
\begin{proof}
The choice of learning rates implies that there is a separation of timescales between the updates of \eqref{eq:q_val_update} and \eqref{eq:thres_update}. In particular, since $b^2_n/b^1_n \rightarrow 0$, iteration \eqref{eq:q_val_update} evolves at a faster timescale than iteration \eqref{eq:thres_update}. Therefore, we first consider update \eqref{eq:thres_update} under the assumption that the policy $\pi_{\tau}$, which updates at the slower timescale, is constant. 

\setlength{\parindent}{0em} We first provide a preliminary result.
\begin{lemma}
\label{lemma:lipscitz}
Let $Q_{\tau}$ denote the action-value function corresponding to the policy $\pi_\tau$. Then, 
$Q_\tau$ is Lipscitz continuous in $\tau$.
\begin{proof}
This follows immediately from the Lipscitz continuity of $\pi_\tau$ in $\tau$.
\end{proof}
\end{lemma}

Define the operator $\mathcal{M}_\tau : \mathds{R}^\mathsf{N} \rightarrow \mathds{R}^\mathsf{N}$, where $\mathsf{N} = (\mathsf{X} + 1) \times (\mathsf{L} + 1) \times \mathcal{A}$, as follows:
\begin{multline}
    [\mathcal{M}_\tau Q](x,\ell,a) = \bigl[\bar\rho(x,\ell,a) + 
    \beta \sum_{x',\ell'} p(x',\ell' | x, \ell, a) \\ \min_{a' \in \mathcal{A}} Q(x',\ell',a')] - Q(x,\ell,a). 
\end{multline}

Then, the step-size conditions on $\{b^1_n\}_{n \ge 1}$ imply that for a fixed $\pi_\tau$, iteration \eqref{eq:q_val_update} may be viewed as a noisy discretization of the ODE (ordinary differential equation):
\begin{equation}
    \dot{Q}(t) = \mathcal{M}_\tau[Q(t)].
    \label{eq:ode_val}
\end{equation}

Then we have the following:
\begin{lemma}
\label{lemma:ode_val}
The ODE \eqref{eq:ode_val} has a unique globally asymptotically stable equilibrium point $Q_{\tau}$.
\begin{proof}
Note that the ODE \eqref{eq:ode_val} may be written as 
\[
    \dot{Q}(t) = \mathcal{B}_\tau[Q(t)] - Q(t)
\]
where the Bellman operator $\mathcal{B}_\tau : \mathds{R}^\mathsf{N} \rightarrow \mathds{R}^\mathsf{N}$ is given by
\begin{multline}
    \mathcal{B}_\tau[Q](x,\ell) = \bigl[\bar\rho(x,\ell,a) +
    \beta \sum_{x',\ell'} p(x',\ell' | x, \ell, a) \\ \times \min_{a' \in \mathcal{A}} Q(x',\ell',a')].
\end{multline}
Note that $\mathcal{B}_\tau$ is a contraction under the sup-norm. Therefore, by Banach fixed point theorem, $Q = \mathcal{B}_\tau Q$ has a unique fixed point, which is equal to $Q_\tau$. The result then follows from ~\cite[Theorem 3.1]{borkar1997analog}.
\end{proof}
\end{lemma}

We now consider the faster timescale. Recall that $(x_0, \ell_0)$ is the initial state of the MDP. Recall
\[
J(\tau) = V_{\tau}(x_0,\ell_0)
\]
and consider the ODE limit of the slower timescale iteration \eqref{eq:thres_update}, which is given by
\begin{equation}
    \dot{\tau} = - \nabla J(\tau).
    \label{eq:ode_thres}
\end{equation}

\begin{lemma}
\label{lemma:ode_thres}
The equilibrium points of the ODE \eqref{eq:ode_thres} are the same as the local optima of $J(\tau)$. Moreover, these equilibrium points are locally asymptotically stable.
\begin{proof}
The equivalence between the stationary points of the ODE and local optima of $J(\tau)$ follows from definition. Now consider $J(\tau(t))$ as a Lyapunov function. Observe that
\[
\frac{d}{dt} J(\tau(t)) = - \bigl[\nabla J(\tau(t))]^2 < 0, 
\]
as long as $\nabla J(\tau(t)) \ne 0$. 
Thus, from Lyapunov stability criteria all local optima of \eqref{eq:ode_thres} are locally asymptotically stable.
\end{proof}
\end{lemma}

Now, we have all the ingredients to prove convergence. Lemmas \ref{lemma:lipscitz}-\ref{lemma:ode_thres} imply assumptions (A1) and (A2) of \cite{Borkar1997}. Thus, the iteration \eqref{eq:q_val_update} and \eqref{eq:thres_update} converges almost surely to a limit point $(Q^{\circ}, \tau^{\circ})$ such that $Q^{\circ} = Q_{\tau^{\circ}}$ and $\nabla J(\tau^{\circ}) = 0$ provided that the iterates $\{Q_n\}_{n \ge 1}$ and  $\{\tau_n\}_{n \ge 1}$ are bounded.

Note that $\{\tau_n\}_{n \ge 1}$ are bounded by construction. The boundness of $\{Q_n\}_{n \ge 1}$ follows from considering the scaled version of \eqref{eq:ode_val}:
\begin{equation}
    \dot{Q} = \mathcal{M}_{\tau,\infty} Q
    \label{eq:ode_scaled}
\end{equation}
where,
\[
\mathcal{M}_{\tau,\infty} Q = \lim_{c \rightarrow \infty} \frac{\mathcal{M}_\tau[c Q]}{c}.
\]

It is easy to see that 
\begin{multline}
    [\mathcal{M}_{\tau,\infty} Q](x, \ell, a) = \beta \sum_{x',\ell'} p(x',\ell' | x, \ell, a) \min_{a' \in \mathcal{A}} Q(x',\ell',a') \\ - Q(x,\ell,a)
\end{multline}

Furthermore, origin is the asymptotically stable equilibrium point of \eqref{eq:ode_scaled}. Thus, from \cite{borkar2000ode}, we get that the iterates $\{Q_n\}_{n \ge 1}$ of \eqref{eq:q_val_update} are bounded.
\end{proof}
\section*{Acknowledgment}

The numerical experiments were enabled in part by support provided by Compute Canada. The authors are grateful to Pierre Thibault from Ericsson Systems for setting up the virtual machines to run the docker-testbed experiments. The authors also acknowledge all the help and support from Ericsson Systems, especially the Global Aritifical Intelligent Accelerator (GAIA) Montreal Team.

\ifCLASSOPTIONcaptionsoff
  \newpage
\fi

\bibliographystyle{IEEEtran}
\bibliography{IEEEabrv,main}

\begin{IEEEbiography}[{\includegraphics[width=1in,height=1.25in,clip,keepaspectratio]{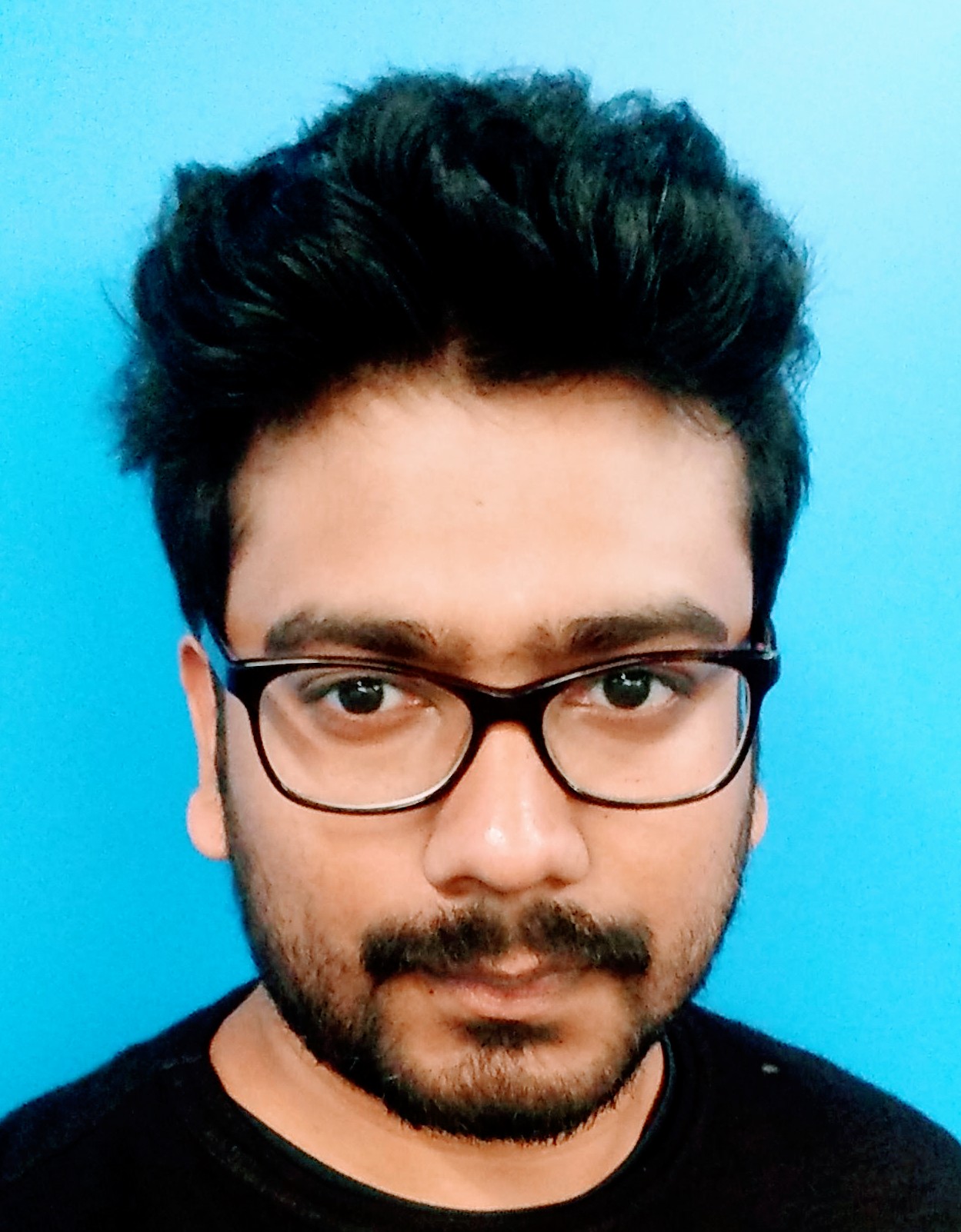}}]{Anirudha Jitani}
received the B.Tech degree in Computer Science and Engineering from Vellore Institute of Technology, India in 2015. He is pursuing his M.Sc. in Computer Science at McGill University, Montreal since 2018. He is a research assistant at Montreal Institute of Learning Algorithms (MILA) and a Data Scientist Intern at Ericsson Systems. He also worked as a software developer for Cisco Systems and Netapp. His research interests include application of machine learning techniques such as deep reinforcement learning, multi-agent reinforcement learning, and graph neural networks in communication and networks.
\end{IEEEbiography}

\begin{IEEEbiography}[{\includegraphics[width=1in,height=1.25in,clip,keepaspectratio]{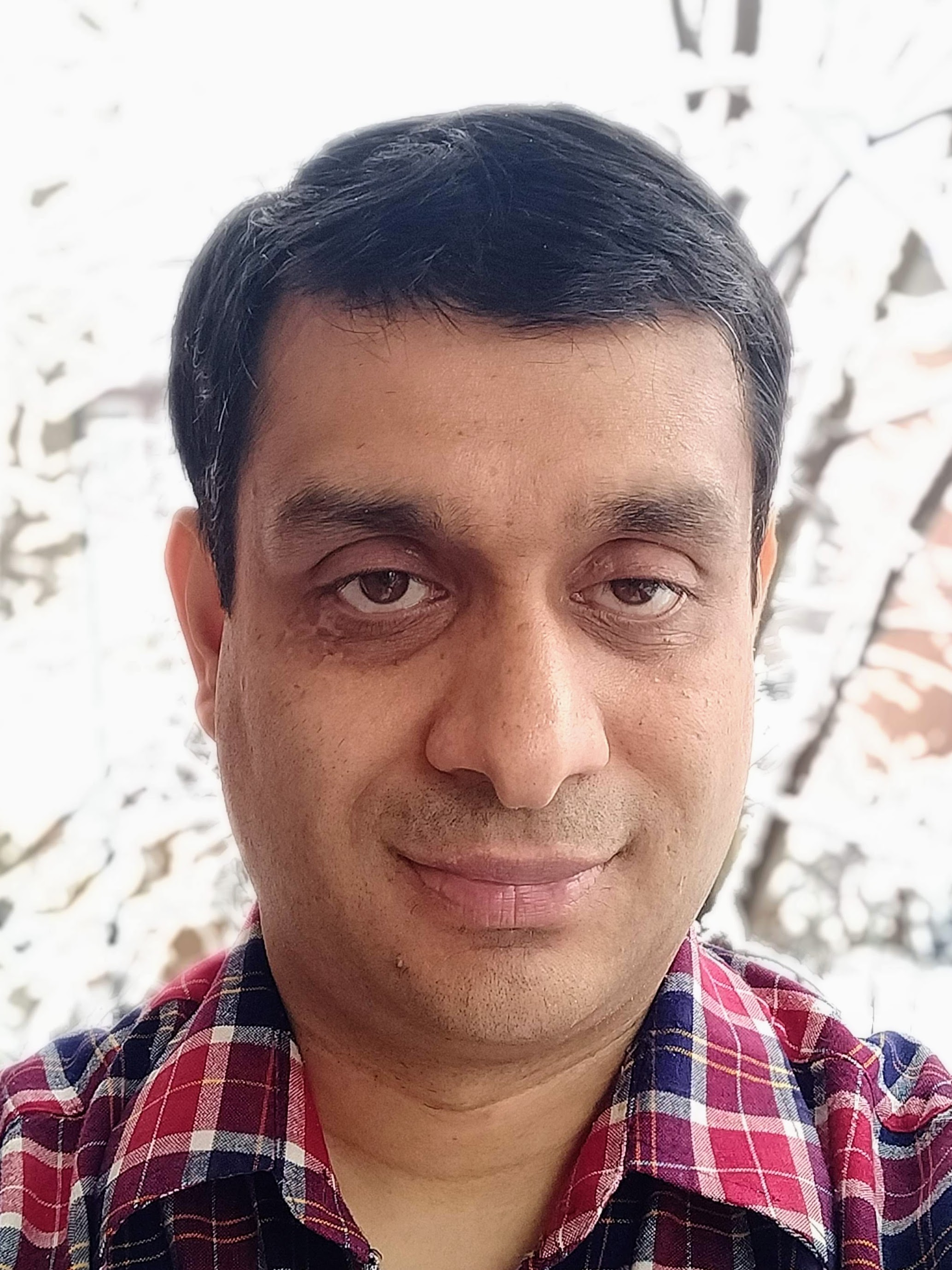}}]{Aditya Mahajan}
(S’06-M’09-SM’14) received
B.Tech degree from the Indian Institute of Technology, Kanpur, India, in 2003, and M.S. and Ph.D.
degrees from the University of Michigan, Ann Arbor,
USA, in 2006 and 2008. From 2008 to 2010, he was
a Postdoctoral Researcher at Yale University, New
Haven, CT, USA. He has been with the department
of Electrical and Computer Engineering, McGill
University, Montreal, Canada, since 2010 where he is
currently Associate Professor. He serves as
Associate Editor of Springer Mathematics of Control,
Signal, and Systems. He was an Associate Editor of the IEEE Control Systems
Society Conference Editorial Board from 2014 to 2017. He is the recipient
of the 2015 George Axelby Outstanding Paper Award, 2014 CDC Best
Student Paper Award (as supervisor), and the 2016 NecSys Best Student Paper
Award (as supervisor). His principal research interests include decentralized
stochastic control, team theory, multi-armed bandits, real-time communication,
information theory, and reinforcement learning.
\end{IEEEbiography}


\begin{IEEEbiography}[{\includegraphics[width=1in,height=1.25in,clip,keepaspectratio]{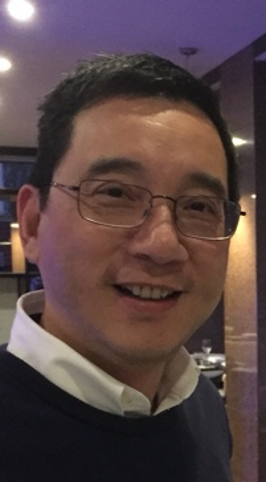}}]{Zhongwen Zhu} received the B.Eng in Mechanical Engineering and M.Eng. in Turbomachinery from Shanghai Jiao Tong University, Shanghai, P.R. China, and also a Ph.D. in Applied Science from Free University of Brussels, Belgium, in 1996. He joined in Aerospace Lab in National Research Council of Canada (NRC) in 1997. One year later, he started to work in Bombardier Aerospace. Since 2001, he has been working for Ericsson Canada with different roles, e.g. Data Scientist, SW Designer, System manager, System designer, System/Solution Architect, Product owner, Product manager, etc. He was an associated editor for Journal of Security and Communication Network (Wiley publisher). He was the invited Technical committee member for International Conference on Multimedia Information Networking and Security. He was the recipient of the best paper award in 2008 IEEE international conference on Signal Processing and Multimedia applications. His current research interests are 5G network, network security, edge computing, Reinforcement learning, Machine learning for 2D/3D object detections, IoT (Ultra Reliable Low Latency application), etc.
\end{IEEEbiography}

\begin{IEEEbiography}[{\includegraphics[width=1in,height=1.25in,clip,keepaspectratio]{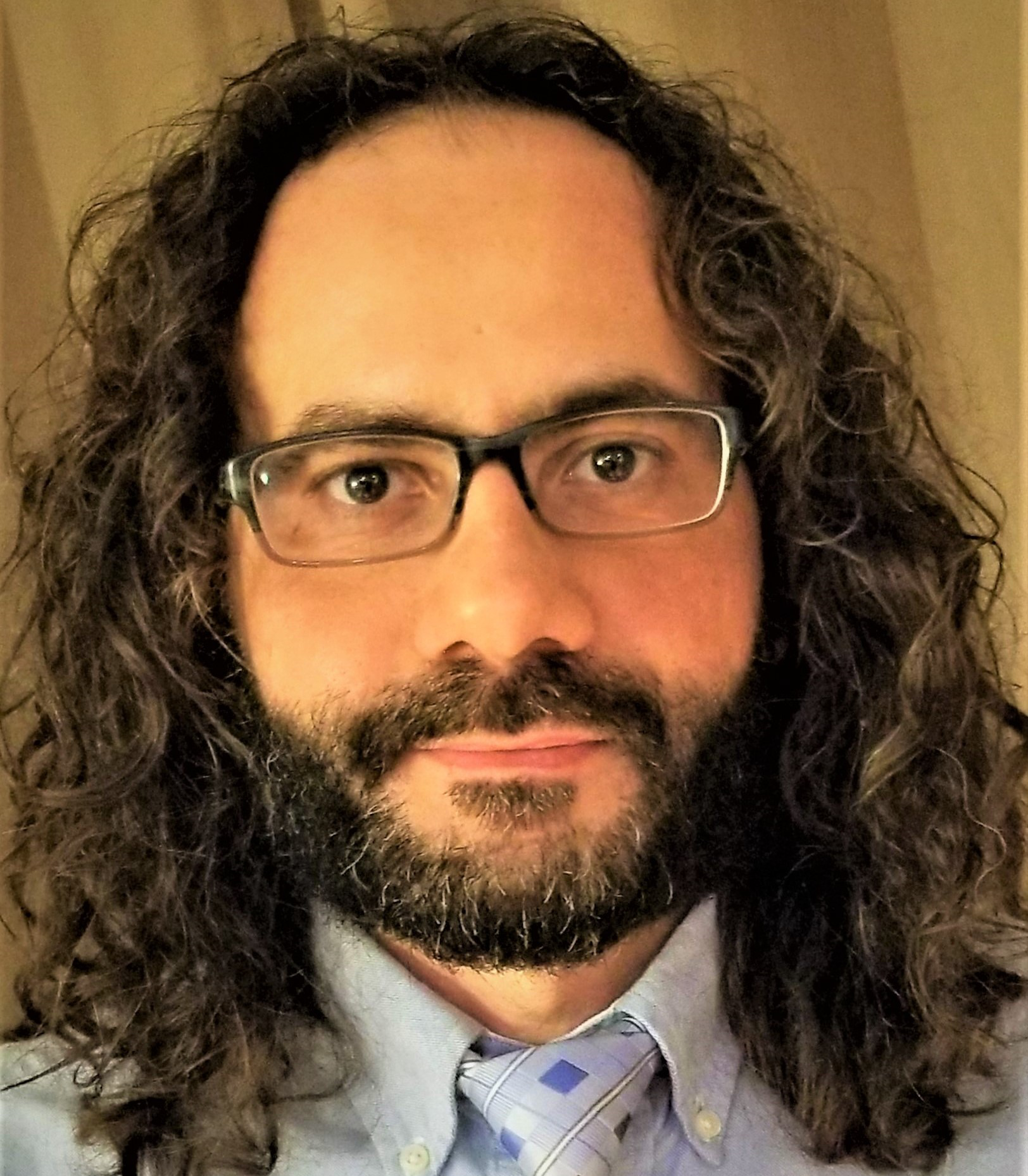}}]{Hatem Abou-Zeid} is a 5G Systems Developer at Ericsson Canada. He has 10+ years of R\&D experience in communication networks spanning radio access networks, routing, and traffic engineering. He currently leads 5G system designs and intellectual property development in the areas of network intelligence and low latency communications. He serves on the Ericsson Government Industry Relations and University Engagements Committee where he directs academic research partnerships on autonomous networking and augmented reality communications. His research investigates the use of reinforcement learning, stochastic optimization, and deep learning to architect robust 5G/6G networks and applications - and his work has resulted in 60+ filed patents and publications in IEEE flagship journals and conferences. Prior to joining Ericsson, he was at Cisco Systems designing scalable traffic engineering and IP routing protocols for service provider and data-center networks. He holds a Ph.D. in Electrical and Computer Engineering from Queen's University. His collaborations with industry through a Bell Labs DAAD RISE Fellowship led to the commercialization of aspects of predictive video streaming, and his Thesis was nominated for an Outstanding Thesis Medal.

\end{IEEEbiography}

\begin{IEEEbiography}[{\includegraphics[width=1in,height=1.25in,clip,keepaspectratio]{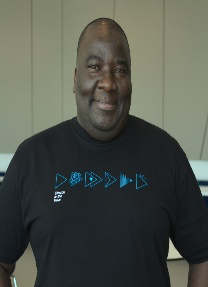}}]{Emmanuel Thepie Fapi} is currently a Data Scientist with Ericsson Montreal, Canada. He holds a master’s degree in engineering mathematics and computer tools from Orleans University in France and a PhD in signal processing and telecommunications from IMT Atlantique in France (former Ecole des Telecommunications de Bretagne) in 2009. From 2010 to 2016 he worked with GENBAND US LLC, QNX software System Limited as audio software developer, MDA system as analyst and EasyG as senior DSP engineer in Vancouver, Canada. In 2017 he joined Amazon Lab 126 in Boston, USA as audio software developer for echo dot 3rd generation. His main areas of interest are 5G network, anomaly detection-based AI/ML, multi-resolution analysis for advanced signal processing, real-time embedded OS and IoT, voice and audio quality enhancement.
\end{IEEEbiography}

\begin{IEEEbiography}[{\includegraphics[width=1in,height=1.25in,clip,keepaspectratio]{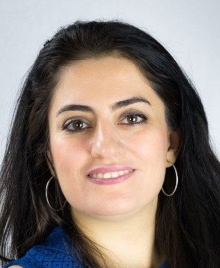}}]{Hakimeh Purmehdi} is a data scientist at Ericsson Global Artificial Intelligence Accelerator, which leads innovative AI/ML solutions for future wireless communication networks. She received her Ph.D. degree in electrical engineering from the Department of Electrical and Computer Engineering, University of Alberta, Edmonton, AB, Canada. After completing a postdoc in AI and image processing at the Radiology Department, University of Alberta, she co-founded Corowave, a startup to develop bio sensors to monitor human vital signals by leveraging radio frequency technology and machine learning. Before joining Ericsson, she was with Microsoft Research (MSR) as a research engineer, and contributed in the development of TextWorld, which is a testbed for reinforcement learning research projects. Her research focus is basically on the intersection of wireless communication (5G and beyond including resource management and edge computing), AI solutions (such as online learning, federated learning, reinforcement learning, deep learning), optimization, and biotech.
\end{IEEEbiography}




\end{document}